\documentclass{jpconf}

\usepackage{hyperref}
\usepackage{longtable}
\usepackage{graphicx}
\usepackage{tabularx}
\usepackage{amsmath,amssymb}
\usepackage{txfonts}
\usepackage{epstopdf}
\usepackage{bm}
\usepackage{xcolor}

\definecolor{violet}{rgb}{0.58,0,0.82}
\definecolor{indigo}{rgb}{0.29,0,0.51}
\definecolor{blue}{rgb}{0,0.5,1}
\definecolor{green}{rgb}{0.46,0.67,0.19}
\definecolor{yellow}{rgb}{0.92,0.69,0.12}
\definecolor{orange}{rgb}{0.85,0.32,0.1}
\definecolor{red}{rgb}{0.64,0.08,0.18}

\newcolumntype{L}[1]{>{\raggedright\arraybackslash}p{#1}}
\newcolumntype{C}[1]{>{\centering\arraybackslash}p{#1}}
\newcolumntype{R}[1]{>{\raggedleft\arraybackslash}p{#1}}
\newcolumntype{Y}{>{\centering\arraybackslash}X}

\newcommand\vv{\bm{v}}
\newcommand\xv{\bm{x}}
\newcommand\Bv{\bm{B}}
\newcommand\Jv{\bm{J}}
\newcommand\Ev{\bm{E}}

\newcommand\uv{\bm{u}}
\newcommand\kv{\bm{k}}
\newcommand\di{d_{\rm i}}

\begin{document}

\title{Three-dimensional simulations of solar wind turbulence with the hybrid code CAMELIA}

\author{L. Franci$^1$, P. Hellinger$^2$, M. Guarrasi$^3$, C. H. K. Chen$^4$, E. Papini$^1$, A. Verdini$^1$, L. Matteini$^{5}$, and S. Landi$^{1,6}$}
\address{$^1$Dipartimento di Fisica e Astronomia, Universit\`a degli Studi di Firenze, Firenze, Italy}
\address{$^2$Astronomical Institute, CAS, Prague, Czech Republic}
\address{$^3$SuperComputing Applications and Innovation Department, CINECA, Bologna, Italy}
\address{$^4$School of Physics and Astronomy, Queen Mary University of London, London, UK}
\address{$^5$LESIA-Observatoire de Paris, Meudon, France}
\address{$^6$INAF - Osservatorio Astrofisico di Arcetri, Firenze, Italy}
\ead{franci@arcetri.astro.it}

\date{\today}
 
\begin{abstract}
We investigate the spectral properties of plasma turbulence from fluid
to sub-ion scales by means of high-resolution three-dimensional
(3D) numerical simulations performed with the hybrid particle-in-cell
(HPIC) code CAMELIA.  We produce extended turbulent spectra with
well-defined power laws for the magnetic, ion bulk velocity, density,
and electric fluctuations. The present results are in good agreement
with previous two-dimensional (2D) HPIC simulations, especially in the
kinetic range of scales, and reproduce several features observed in
solar wind spectra. By providing scaling tests on many different
architectures and convergence studies, we prove CAMELIA to represent a
very efficient, accurate and reliable tool for investigating the
develpoment of the turbulent cascade in the solar wind, being able to
cover simultaneously several decades in wavenumber, also in 3D.
\end{abstract}

\section{Introduction}
\label{sec:introduction}

Turbulence in magnetized collisionless plasmas, such as the solar
wind, is one of the major challenges of space physics and
astrophysics. Both the anisotropic flow of energy toward smaller
scales (cascade) and the damping of energy at dissipative scales are
poorly understood.  Solar wind turbulent fluctuations are generated at
large magneto-hydrodynamic (MHD) scales and dissipated at scales where
particle kinetics dominates. Indeed, in situ measurements show spectra
of the plasma and electromagnetic fields with a power-law scaling
spanning several decades in frequency, with a spectral break in the
magnetic and density spectrum at proton scales, separating the MHD
inertial range cascade from a second power-law interval at kinetic
scales (see~\cite{Chen_2016} for a recent summary).
A further change in the spectral properties occurs
at the electron scales, although distinguishing between an exponential 
cut-off \cite{Alexandrova_al_2012} or a power law \cite{Sahraoui_al_2013} 
is not straightforward and thus no universal behavior is observed.

Very large numerical resources are required to investigate
the whole turbulent cascade, since at least two full decades in
wavevectors across the transition needs to be covered
simultaneously. Full Particle-In-Cell (PIC) simulations represent the
most comprehensive numerical tool for simulating the plasma dynamics
up to electron spatial and temporal scales, e.g.,
~\cite{Svidzinski_al_2009, Saito_al_2010, Chang_al_2011,
  Camporeale_Burgess_2011, Gary_al_2012, Wan_al_2012,
  Karimabadi_al_2013, Wan_al_2015, Gary_al_2016,
  Wan_al_2016}. However, due to computational limitations, they
typically employ a limited accuracy (e.g., small resolution, small
number of particles, small ion-to-electron mass ratio).

Alternatively, reduced models have been largely used in the last
decade to simulate plasma turbulence at kinetic scales, both in 2D and
3D, e.g., Hall-MHD~\cite{Shaikh_Shukla_2009, Shaikh_Zank_2009,
  Martin_al_2013, Rodriguez_al_2013, Gomez_al_2013},
Electron-MHD~\cite{Cho_Lazarian_2009, Shaikh_2009,
  Meyrand_Galtier_2013}, Gyrokinetic~\cite{Howes_al_2011,
  TenBarge_al_2013a, Hatch_al_2014, Li_al_2016}, finite Larmor radius-Landau
fluid~\cite{Passot_al_2014, Sulem_Passot_2015, Kobayashi_al_2017},
hybrid Vlasov-Maxwell~\cite{Servidio_al_2012, Greco_al_2012,
  Perrone_al_2014b, Valentini_al_2014, Servidio_al_2015,
  Valentini_al_2017, Cerri_Califano_2016, Cerri_al_2017}, and Hybrid
Particle-In-Cell (HPIC) \cite{Verscharen_al_2012,
  Vasquez_Markovskii_2012, Comisel_al_2013, Parashar_al_2014,
  Vasquez_al_2014, Hellinger_al_2015, Ozak_al_2015, Parashar_al_2015,
  Parashar_Matthaeus_2016}.

In particular, very high-resolution 2D simulations, performed with the
HPIC code CAMELIA, recently fully covered the transition between fluid
and kinetic scales \cite{Franci_al_2015a, Franci_al_2015b,
  Franci_al_2016b}. Moreover, they produced extended turbulent spectra with
well-defined power laws for the plasma and electromagnetic fields, in
agreement with solar wind observations.  Such results have been
recently extended to 3D~\cite{Franci_al_2017b}.

\section{Numerical setup}
\label{sec:setup}

The simulations presented here have been performed using the
3D HPIC code CAMELIA (see Sec.~\ref{sec:camelia} for details).  
The characteristic spatial
unit is the ion (proton) inertial length, $\di=v_A/\Omega_{\rm i}$,
$v_A$ being the Alfv\'en speed, while the temporal unit is the inverse
ion gyrofrequency, $1/\Omega_{\rm i}$. We compare four 
simulations, all employing a periodic cubic grid with spatial
resolution $\Delta x = 0.25 \, \di$, with two different numbers of grid
points ($512^3$ and $256^3$) and, consequently, two different box
sizes ($L_{\rm box} = 128\,\di$ and $64 \di$), and with three different
numbers of (macro)particle-per-cell, ppc, representing protons ($512$,
$1024$, and $2048$). In all four cases, the equilibrium system is a homogeneous plasma,
with uniform density and temperature, embedded in a mean
magnetic field, $\Bv_0$. The plasma beta, i.e., the ratio of the plasma pressure
to the magnetic pressure, is $\beta = 0.5$ for both the protons and the electrons.
This system is perturbed with 
magnetic and ion bulk velocity fluctuations, purely perpendicular to $\Bv_0$,
consisting of a 
superposition of Fourier modes of equal amplitude and random phases in
the range $k_0 < k  < k_{\mathrm{inj}}$ ($k =
\sqrt{\kv_x^2 + \kv_y^2+ \kv_z^2}$). The minimum wavenumber, $k_0 = 2
\pi / L_{\rm box}$, is $0.05 \, \di^{-1}$ for the large box 
and $0.10 \, \di^{-1}$ for the small one, while the maximum
injection scale is $k_{\mathrm{inj}} = 0.25 \, \di^{-1}$ and
$0.30 \, \di^{-1}$, respectively. 
A more detailed description of the initializazion and of the
physical and numerical parameters can be found in~\cite{Franci_al_2015a, 
Franci_al_2015b}.
The main differences between the four 3D simulations 
are summarized in Tab.~\ref{tab:parameters} (runs A-D), where we also recall
the parameters of a 2D HPIC simulation with the same $\beta$ (run E), 
presented in \cite{Franci_al_2015a,Franci_al_2015b}.

\begin{center}
\begin{table}[t]
\centering
\begin{tabularx}{0.8\textwidth}{YYYYYYYY}
\br
Run & Grid & $\Delta x$ & $L \,[\di]$ & $\Bv^{\textrm{rms}}/B_0$ &
$k_0 \di$ & $k_{\textrm{inj}} \di$ & ppc \\ 
\mr
A & $512^3$  & $0.25$  & $128$ & $0.40$ & $0.05$  & $0.25$ & $2048$ \\ 
B & $256^3$  & $0.25$  & $64$  & $0.38$ & $0.10$  & $0.30$ & $2048$ \\ 
C & $256^3$  & $0.25$  & $64$  & $0.38$ & $0.10$  & $0.30$ & $1024$ \\ 
D & $256^3$  & $0.25$  & $64$  & $0.38$ & $0.10$  & $0.30$ & $ 512$ \\ 
E & $2048^2$ & $0.125$ & $256$ & $0.24$ & $0.025$ & $0.28$ & $8000$ \\ 
\br
\end{tabularx}
\caption{List of simulations and their main different parameters}
\label{tab:parameters}
\end{table}
\end{center}

\section{Results}
\label{sec:results}

\begin{figure}[t]
\begin{center}
\includegraphics[width=0.85\linewidth]{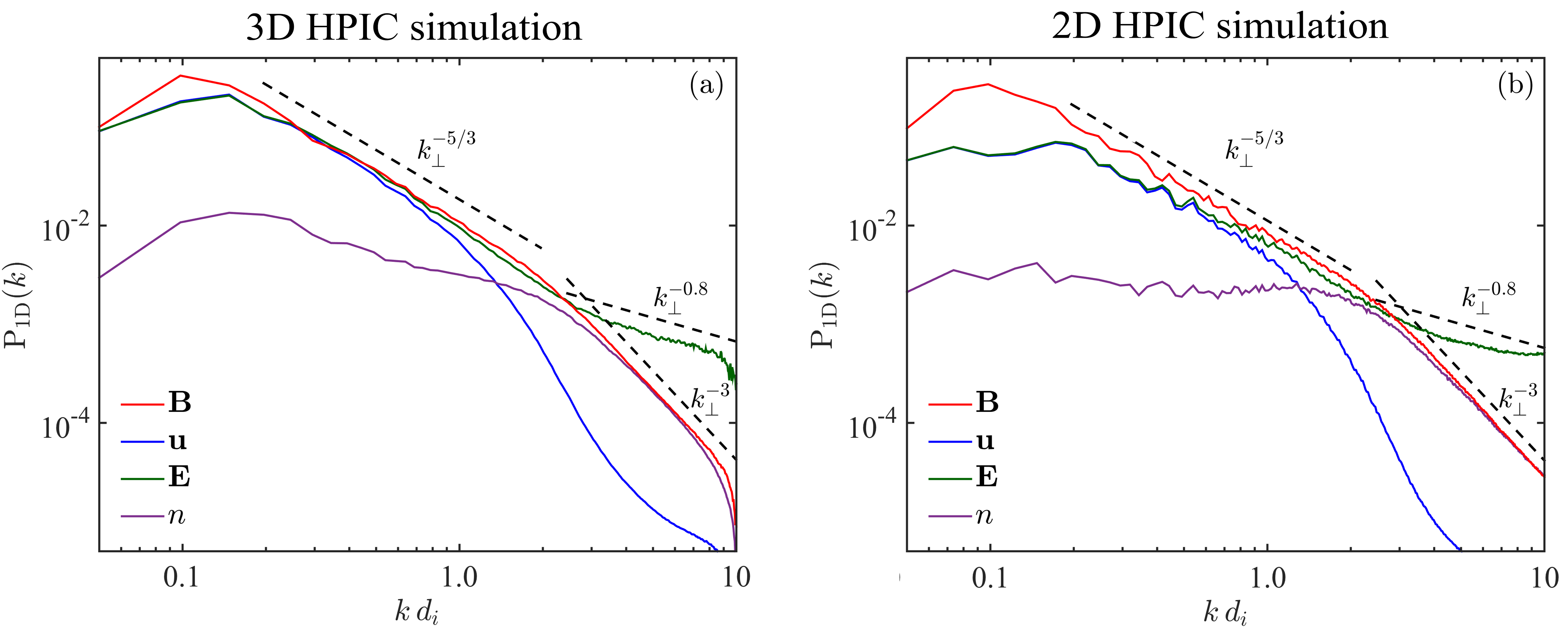}\\
\includegraphics[width=\linewidth]{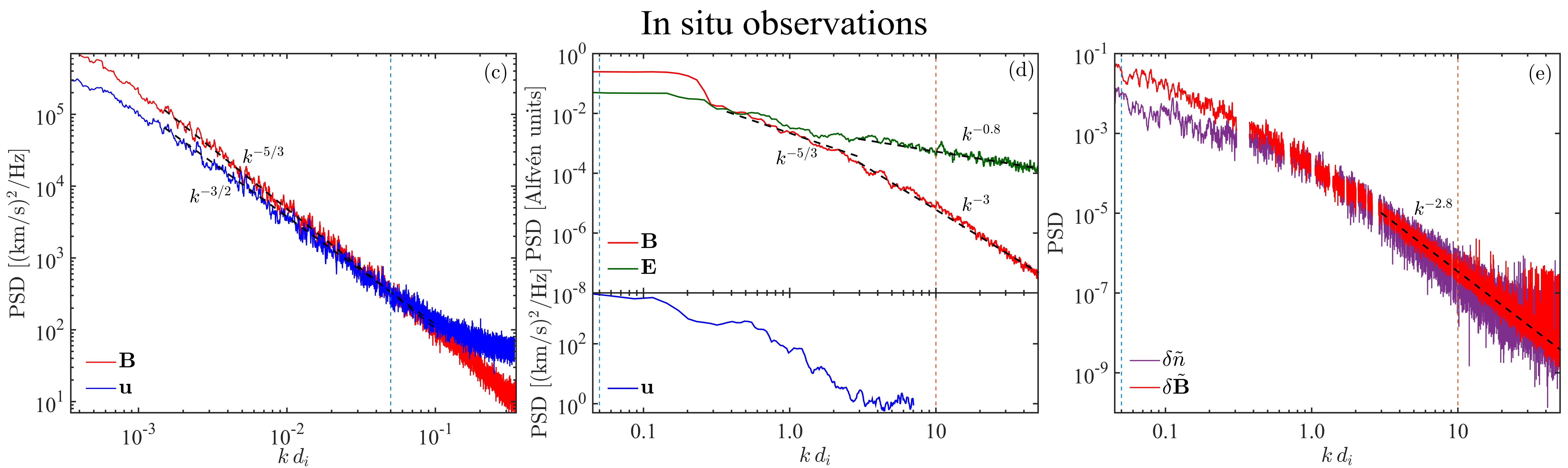}
\caption{Qualitative comparison of the 1D omnidirectional spectra of
  the magnetic field (red), ion bulk velocity (blue), electric field
  (green), and density (purple) between the 3D run A and the 2D run E
  (panels (a) and (b), respectively).  Additionally, the bottom panels
  show results of solar wind and magnetosheath observations
  from~\cite{Chen_al_2013b} (c), \cite{Chen_Boldyrev_2017} (d),
  and~\cite{Chen_al_2013a} (e). A light blue and an orange dashed
  lines mark the smaller and larger wavenumbers covered by the 3D HPIC
  simulation, respectively.}
\label{fig:spectra1DOverview}
\end{center}
\end{figure}
We start with the results of run A, already presented
in~\cite{Franci_al_2017b}, focusing on the spectral properties of
th electromagnetic and plasma fluctuations at the time when the turbulent
cascade has fully developed. 
For the definitions of the 1D omnidirectional spectra, 
$P_{\mathrm{1D}}$, the
1D reduced perpendicular and parallel spectra with respect
to the global mean field $\Bv_0$, 
$P_{\mathrm{1D},\bot}$ and $P_{\mathrm{1D},\parallel}$, 
and for the description of
the filtering procedure used to obtain them from the
3D power spectra, please refer to \cite{Franci_al_2017b}.

In the top panels of Fig.~\ref{fig:spectra1DOverview}, we show a
comprehensive overview of the 1D omnidirectional power spectra of all
fields for the 3D run A (left), to be compared with the results of the
2D run E~\cite{Franci_al_2015a, Franci_al_2015b} (right).
Additionally, characteristic power laws are drawn with dashed black
lines as reference. The main similarities between the 3D and the 2D
cases are: (i) the double power-law behavior of the magnetic
fluctuations, with a spectral index close to $-5/3$ at MHD scales and
a steepening ($\sim -3$) at sub-ion scales, (ii) the strong coupling 
between magnetic and density fluctuations at
sub-ion scales, with the same amplitude and similar slopes,
(iii) the flattening of the electric field
spectrum in the kinetic range, with a spectral index close to $-0.8$,
(iv) the sharp decline of the ion bulk velocity spectrum, which 
quickly reaches the noise level.
All these features are in broad agreement with solar wind observations
\cite{Chen_al_2013b, Chen_Boldyrev_2017, Chen_al_2013a}, as shown in the
bottom panels of Fig.~\ref{fig:spectra1DOverview}, where a light blue and 
a orange dashed lines mark the smaller and larger wavenumbers covered
by the 3D HPIC simulation, respectively.
On the contrary, differences arise at MHD
scales for the velocity, electric field, and density spectra. In
particular, in the 3D case the magnetic and velocity fluctuations 
are more strongly coupled, since they are almost equal amplitude, 
in Alfvén units, up to $k \di \lesssim 1$.
A higher level of density fluctuations is also observed.
Unexpectedly, a better agreement is recovered
between observations and the 2D simulation for what concerns the ion
bulk velocity spectrum. The latter exhibits a lower level of
fluctuations than the 3D case, and therefore a sizeable residual
energy (difference between magnetic and kinetic energy). This
discrepancy between 2D and 3D might be due to the
different geometry or to the different setting, e.g., to the initial
level of fluctuations or resolution (see Tab.~\ref{tab:parameters}).
Indeed, the 3D physical domain is a factor of 2 smaller than the 2D
one and this, and this might constrain the dynamics at the largest
scales.

\begin{figure}[t]
\begin{center}
\includegraphics[width=0.48\linewidth]{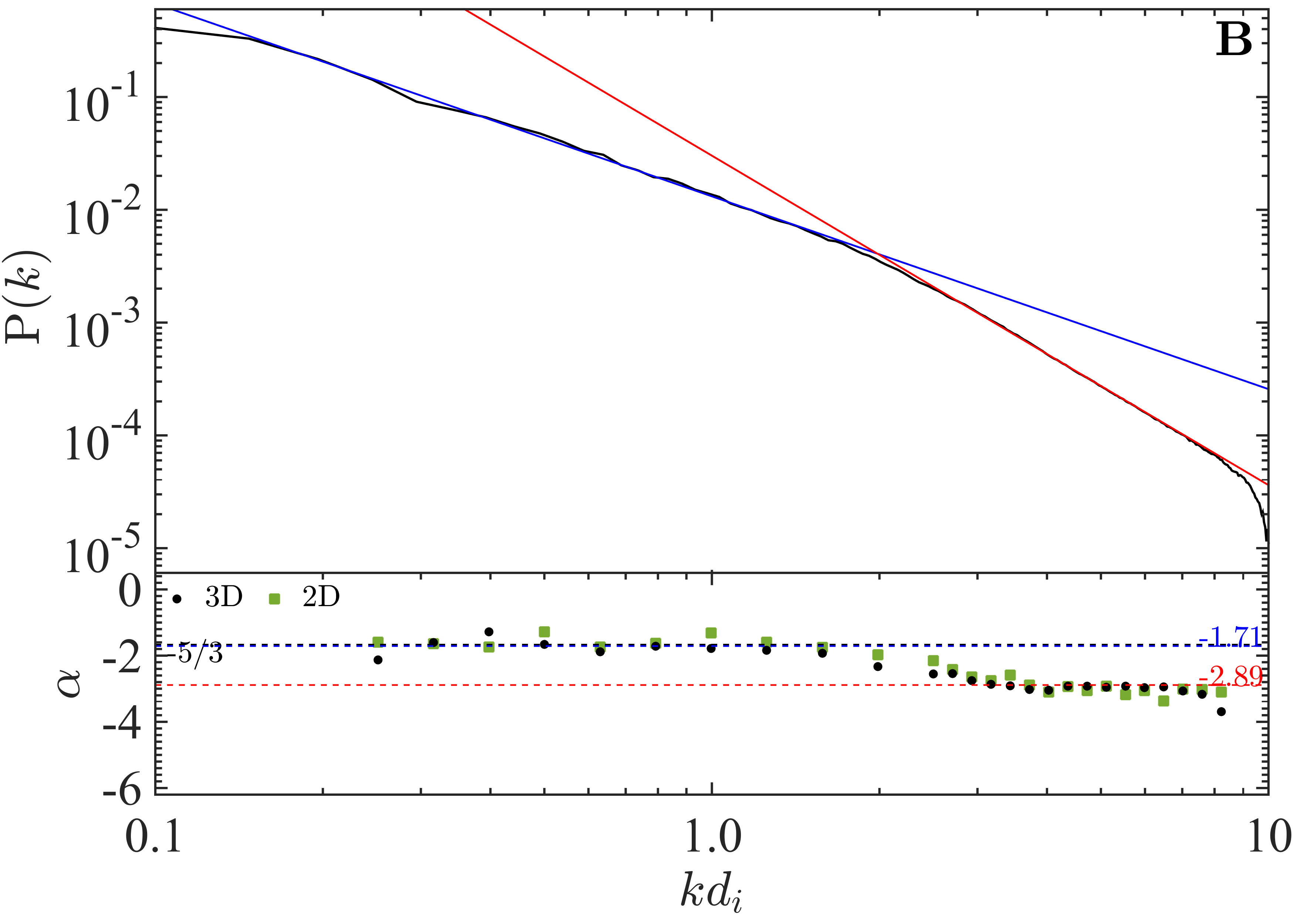}
\includegraphics[width=0.48\linewidth]{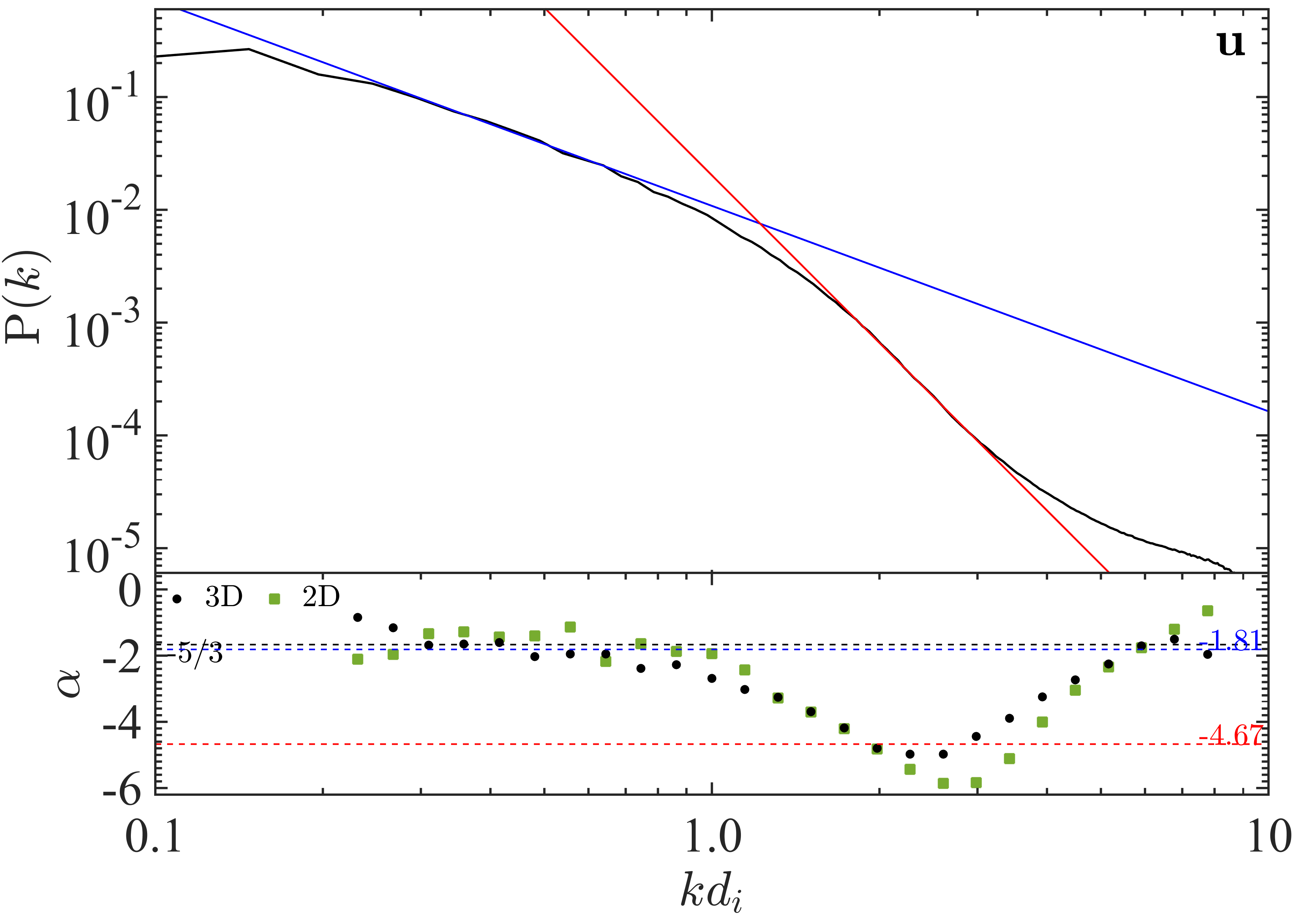}\\
\includegraphics[width=0.48\linewidth]{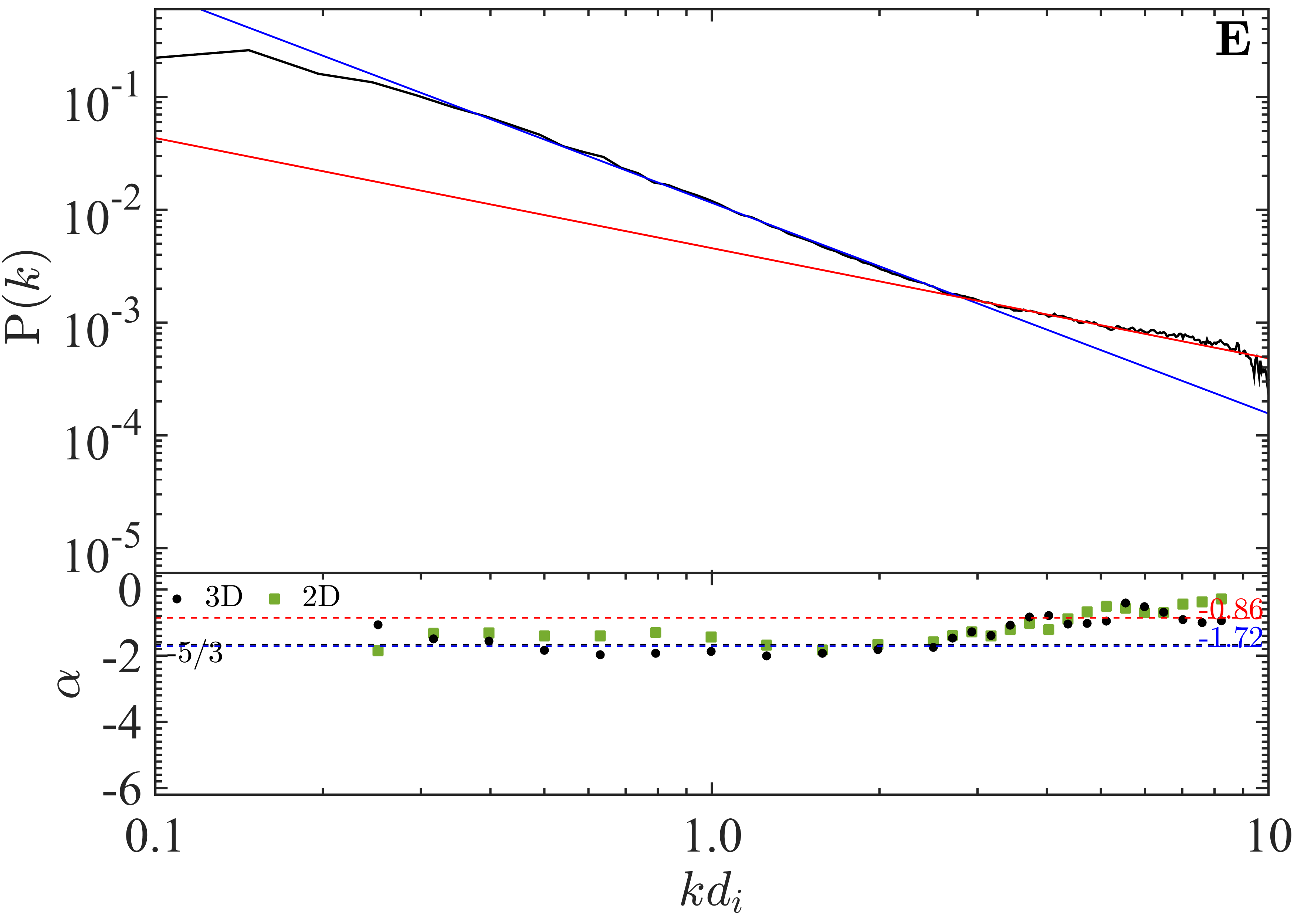}
\includegraphics[width=0.48\linewidth]{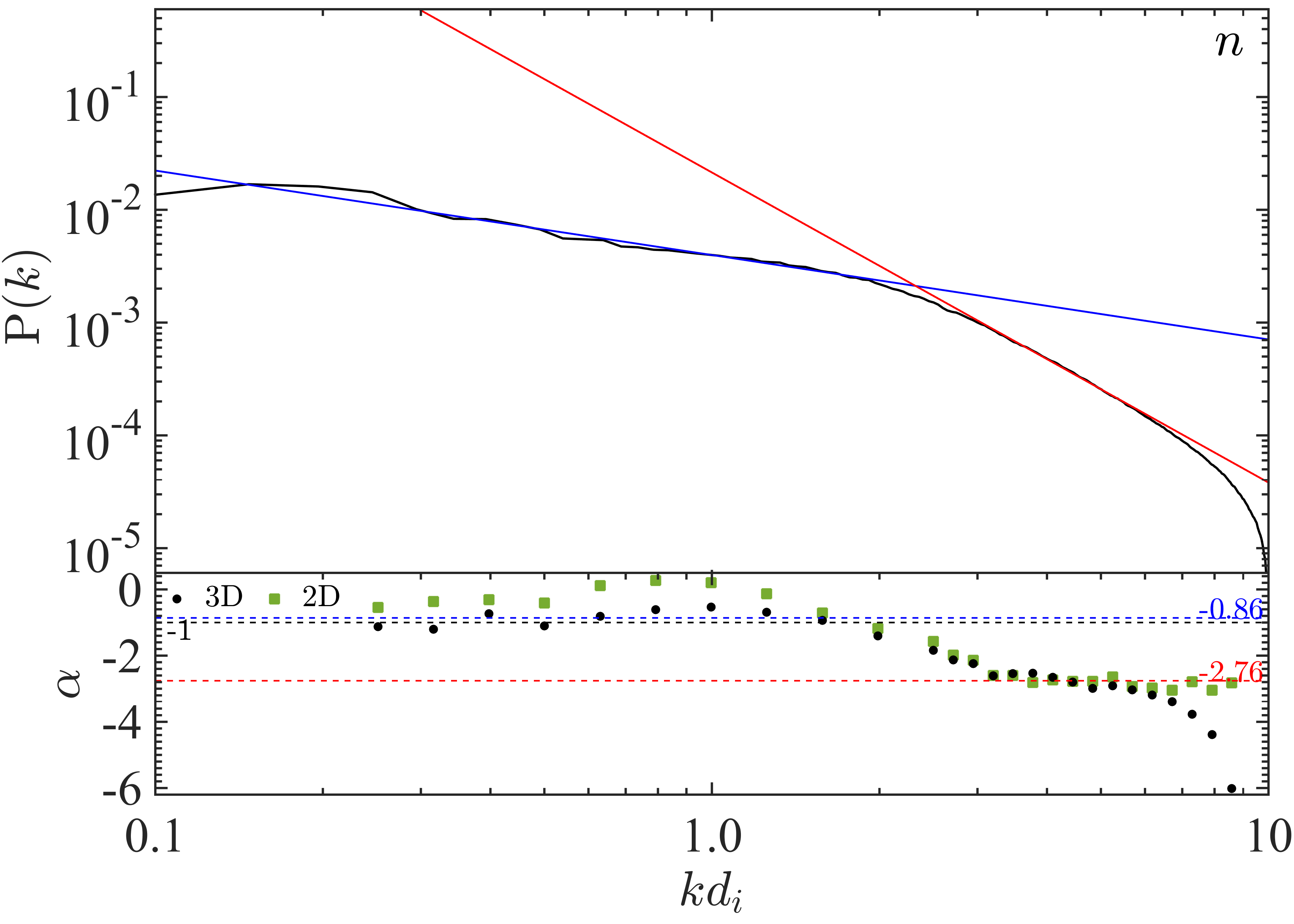}
\caption{1D omnidirectional power spectra of the magnetic field (top
  left), ion bulk velocity (top right), electric field (bottom left),
  and density (bottom right) for run A. Power-law global fits are reported for
  the MHD range (blue) and the kinetic range (red). In the bottom
  region of each panel, the value of the local spectral index,
  $\alpha$ (black), is compared with the results of the two
  global fits in corresponding colors and with the local spectral
  index of the 2D run E (green).}
\label{fig:fitted_spectra}
\end{center}
\end{figure}
In Fig.~\ref{fig:fitted_spectra}, we show in separate panels the 1D 
omnidirectional spectra of the magnetic field, $\Bv$ (top left), ion bulk
velocity, $\uv$ (top right), electric field, $\Ev$ (bottom left), and
density, $n$ (bottom right) for run A. We fit each spectrum with two
power laws, one the MHD range (blu straight line) and one in the kinetic 
range (red), as already done for 2D HPIC simulations~\cite{Franci_al_2016b}.  
Additionally, we perform local
power-law fits over many small intervals in the range
$k_\perp d_i \in [0.25,\,10]$. The values of the local spectral index, 
$\alpha$, are shown with black dots in the
bottom part of each panel. Two horizontal dashed line represent the slopes of
the two global fits, with the respective colors. We also directly 
compare $\alpha$ for the 3D run A (black) and the 2D run E (green).

The magnetic field spectrum exhibits two clear power-law intervals spanning
almost two decades in wavevector, with a spectral index of $\sim-5/3$
and $\sim-2.9$ at MHD and sub-ion scales, respectively, and a
transition at $k \di \gtrsim 2$.  The velocity spectrum shows a
power-law-like behaviour at large scales, although less extended than
the magnetic field's, with a spectral index close to $-5/3$ in 3D
and slightly closer to $-3/2$ in 2D.
For $k \di \gtrsim 1$, the spectrum drops very
rapidly, reaching the ppc noise level at slightly smaller scales. The
hint of a power-law shape can be inferred, 
significantly steeper than the magnetic field's. 
However, the small extent in $k$ prevents us from providing either a clear
evidence of a power law with respect to an exponential drop, 
nor a common value of the slope between the 3D and the 2D runs. 
The electric field spectrum shows an extended
Kolmogorov-like power law at MHD scales and flattens toward a spectral
index of $\sim -0.8$ around ion scales, consistently with the 
generalized Ohm's law \cite{Franci_al_2015b}. Finally, the
density spectrum is almost flat at intermediate scales, with a slope
$\sim -0.9$, although it seems to be slightly steeper at the largest
scales.  A transition is clearly observed around ion scales, followed
by a power law with a spectral index of $\sim -2.7$.

\section{Role of numerical parameters}
\label{sec:convergence}

A very high accuracy is required to investigate the turbulent cascade
from large MHD scales to small kinetic ones. In particular, in order
to quantitatively compare numerical results with solar wind
observations (e.g., the spectral indices of electromagnetic and plasma
fluctuations and the scale of the MHD-kinetic spectral break) one
needs to cover simultaneously at least two full decades in wavevectors
across the transition. Moreover, artificial effects due to the finite
spatial resolution, the finite box size and the finite number of
particles have to be carefully checked, so that the numerical results
can be considered robust and reliable. While all this has been
feasible in 2D in recent years \cite{Franci_al_2015b}, it is not
trivial in 3D. Here we provide such an analysis,  in
support of the results recently presented in \cite{Franci_al_2017b}.

\begin{figure}[t]
\begin{center}
\includegraphics[width=0.45\linewidth]{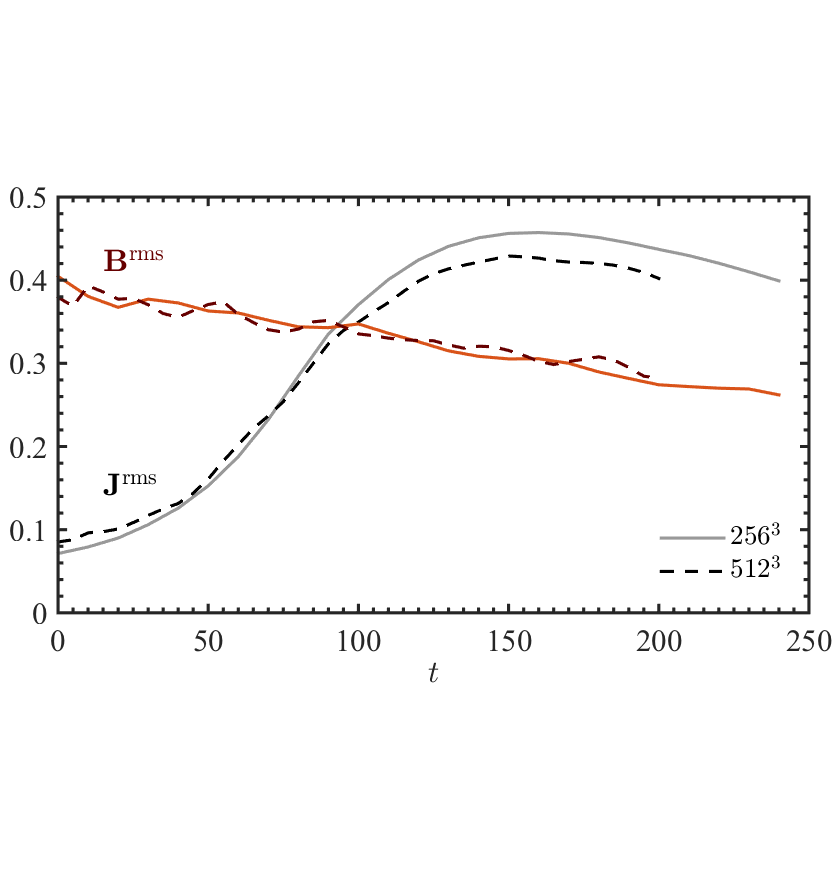}
\includegraphics[width=0.45\linewidth]{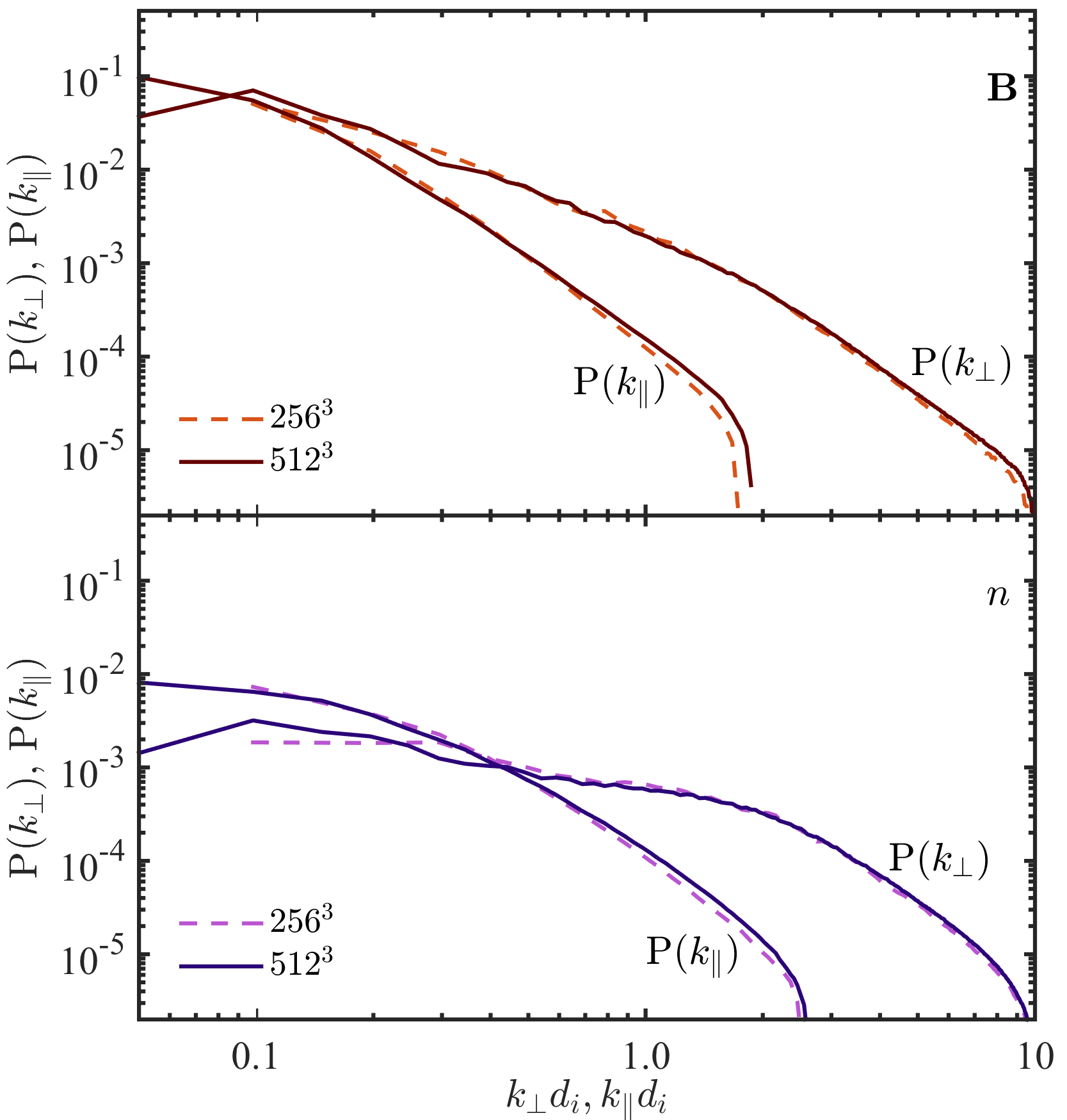}
\caption{Comparison between run A and B, employing different numbers
  of grid points. Left panel: time evolution of rms of the magnetic
  field, $B{\textrm{rms}}$ and of current density,
  $J^{\textrm{rms}}$.  Right panels: 1D reduced perpendicular and
  parallel spectra of the magnetic (red, top) and density (purple,
  bottom) fluctuations.}
\label{fig:convergence_box}
\end{center}
\end{figure}

\begin{figure}[t]
\begin{center}
\includegraphics[width=0.45\linewidth]{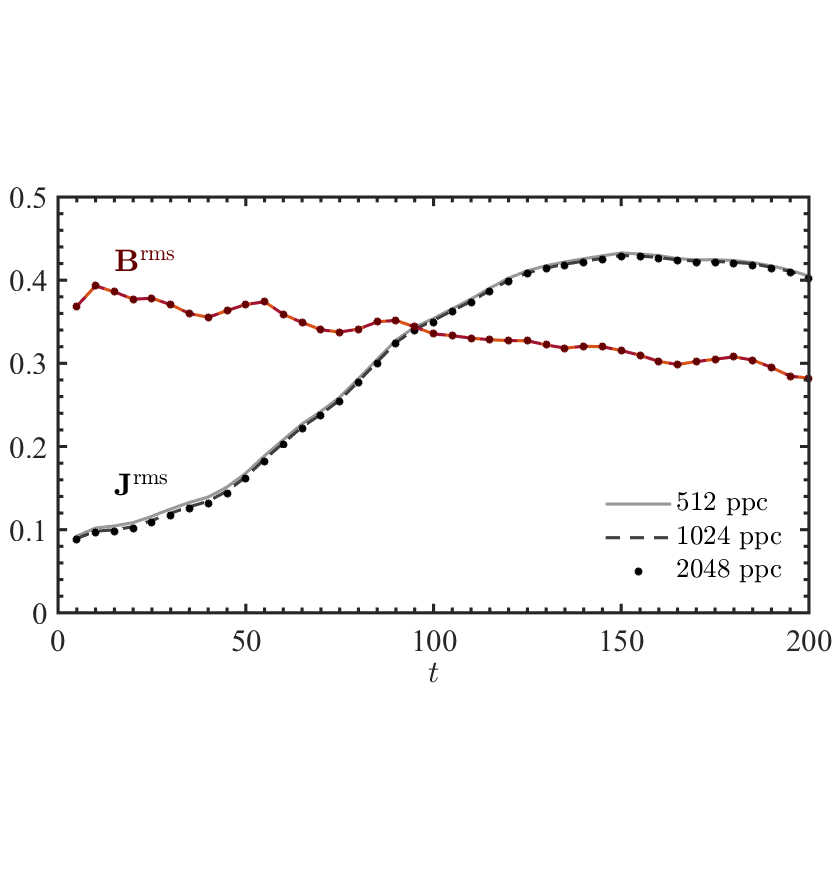}
\includegraphics[width=0.45\linewidth]{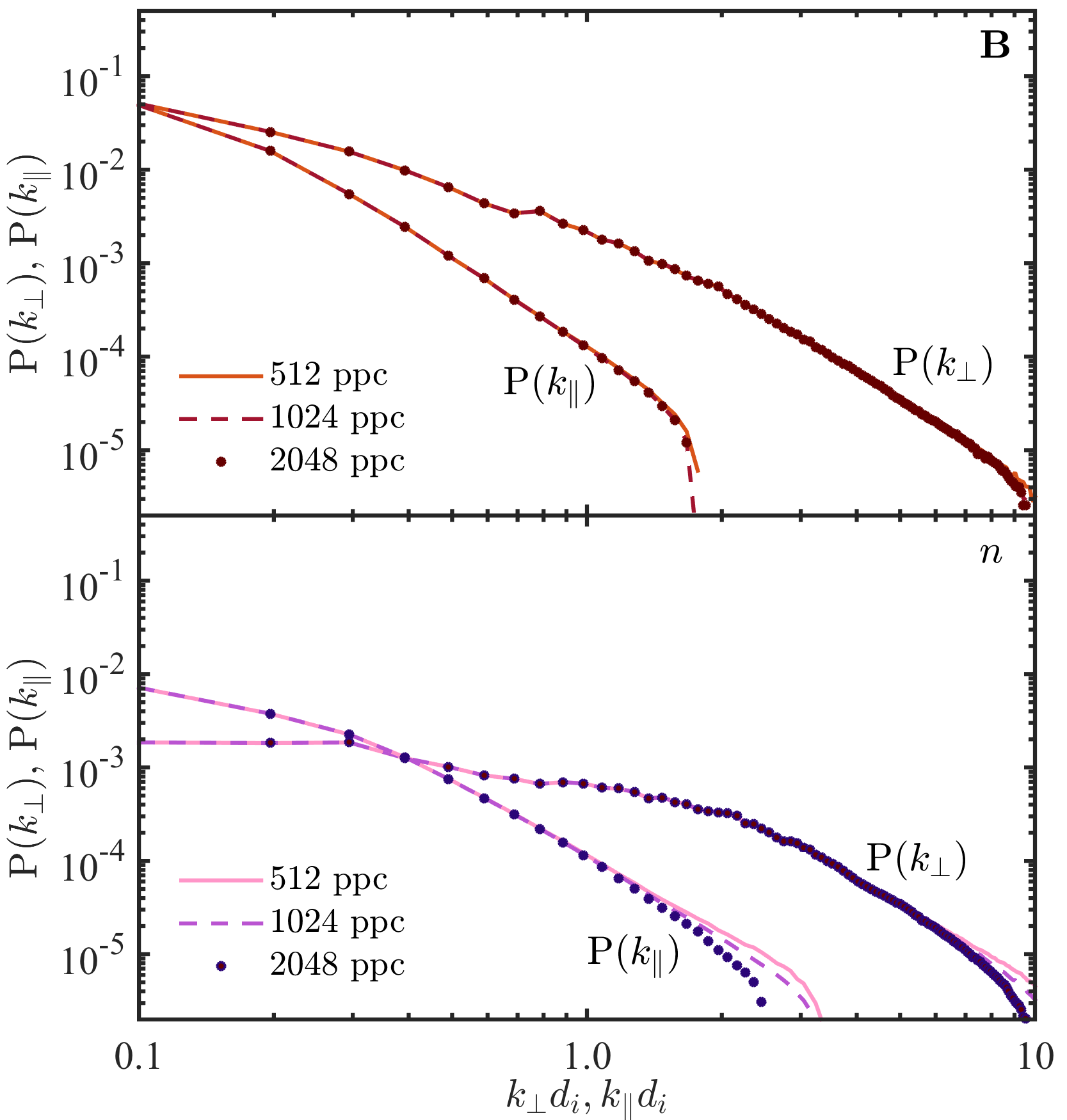}
\caption{The same as in Fig.~\ref{fig:convergence_box}, but for the
comparison between run B, C, and D, employing different numbers
of ppc.}
\label{fig:convergence_ppc}
\end{center}
\end{figure}
 
\subsection{Box size}
\label{subsec:boxsize}

We investigate the effects of the box size by running two simulations
with the same setting and numerical parameters, except for different
grid sizes, $512^3$ (run A) and $256^3$ (run B). Consequently, the box
size changes from $128 \, \di$ to $64 \, \di$. The left panel of
Fig.~\ref{fig:convergence_box} shows that the use of a smaller box
does not affect considerably the time evolution of the system. The rms
value of the magnetic fluctuations, $B^{\mathrm{rms}}$ (red line),
exhibits very small variations around the same average value, with the
same slow decrease, while the rms of the current density,
$J^{\mathrm{rms}}$ (black), shows the very same behavior up to $t
\sim 100$ but then it reaches a slightly smaller peak value.  This
difference is mainly due to the fact that for run B the same
resistivity as for run A was set, while it should be
slightly decreased in order to take into account the small increase in
the mininum wavenumber and in the injection scale.  In the right panel
of Fig.~\ref{fig:convergence_box}, we also compare the reduced
perpendicular and parallel spectra of the total magnetic flucuations
(top) and the density fluctuations (bottom) between runs A and B.  The
differences in $P_\mathrm{1D,\perp}^{\Bv}$ and $P_\mathrm{1D,\perp}^n$
are indeed negligible. Even $P_\mathrm{1D,\parallel}^{\Bv}$ and
$P_\mathrm{1D,\parallel}^n$ show no significant differences, except
for a very small deviation at the smallest parallel scales, that are
not suppressed by the filtering procedure, i.e., for $k_\parallel \,
\di \gtrsim 1$.

\subsection{Number of particles (ppc)}
\label{subsec:particles}

We investigate the effects of the number of particles, in order to
determine the optimal number which allows for obtaining reliable
results at sub-ion scales. We use two simulations employing the small box
and the same parameters as in run B, except for different numbers of ppc, i.e.,
$1024$ (run C) and $512$ (run D) instead of $2048$.  The left panel of
Fig.~\ref{fig:convergence_ppc} shows that the use of fewer particles,
at least a factor of 4, has no effect on the evolution
of global quantities: both $B^{\mathrm{rms}}$ and
$J^{\mathrm{rms}}$ exhibit the same behavior over time.  
The same consideration holds
for the spectral properties of the magnetic fluctuations (top right panel
of Fig.~\ref{fig:convergence_ppc}).  The only, almost negligible,
differences arise in $P_\mathrm{1D,\perp}^n$ (bottom right) at the scales
corresponding to the grid spacing and, correspondingly, in
$P_\mathrm{1D,\parallel}^n$ at scales $k_\parallel \, \di \gtrsim 1$.
Note that the filtering procedure already removes most of the
artificial features due to the noise, making the use of a very large
number of ppc unnecessary. As a consequence, for an intermediate-beta
case such as the one presented here, $512$ ppc could be already 
enough in order to investigate the development of the kinetic cascade
and the spectral properties.  However, if one intends to investigate
the evolution of the proton temperature (not discussed here), the use
of a larger number of ppc would likely be advisable, as previously
shown in 2D~\cite{Franci_al_2015b}.

\section{The hybrid particle-in-cell (HPIC) code CAMELIA}
\label{sec:camelia}

\subsection{Overview and applications}

CAMELIA\footnote{Official CAMELIA website:
  http://terezka.asu.cas.cz/helinger/camelia.html} (Current Advance
Method Et cycLIc leApfrog) is a HPIC code, where the electrons are
considered as a massless, charge neutralizing fluid, whereas the ions
are described as macroparticles, i.e., statistically-representative
portions of the distribution function in the phase space.  Up to 2016,
the code was only parallelized through the Message Passing Interface
(MPI) Library using spatial and/or particle decompositions, with the
former being available in all spatial directions. In early 2017, a
hybrid parallelization using both MPI and the Open MultiProcessing
(OpenMP) Interface has been implemented, which fully exploits the
latest architectures for High Performance Computing, including
many-integrated cores Intel Knights Landing processors.  Different
formats can be chosen for the output files, i.e., ASCII, binary, or
HDF5, including parallel binary (MPI-IO) and parallel HDF5.  A
checkpoint/restart procedure is implemented, with a HDF5 file written
by each MPI process, assuring a wide portability across different
platforms.  The boundary conditions are periodic in all directions.
Alternatively, in 2D, one can choose reflecting boundary conditions on
one side and open boundary conditions, with a continuous injection of
particles, on the other side (e.g., allowing for the generation of
shock waves).

CAMELIA has two major extensions, being able to model (i) the effects
of a slow expansion, using the Hybrid Expanding-Box (HEB) model
(Hellinger et al., 2003; Hellinger and Travnicek, 2005), and (ii) the
effects of Coulomb collisions, using the Langevin representation
(Hellinger and Travnicek, 2010, 2015).  In the last decade, it has
been extensively used to investigate the spectral and heating
properties of solar wind turbulence from MHD to sub-ion scales in
2D~\cite{Franci_al_2015a, Franci_al_2015b}, recovering a good
agreement with solar wind observations. Such findings have also been
validated by comparing with the results obtained by numerical 
simulations performed with the Eulerian hybrid Vlasov-Maxwell code HVM
~\cite{Cerri_al_2017a}. A particular focus has been devoted to 
the correlation between vorticity and proton
temperature~\cite{Franci_al_2016a}, to the dependence of the
ion-scale spectral break on the plasma beta ~\cite{Franci_al_2016b},
and to the role of magnetic reconnection as a trigger for the sub-ion
scale cascade~\cite{Franci_al_2017a}.  The study of the spectral
behavior of electromagnetic and plasma fluctuations has recently been
extended to 3D~\cite{Franci_al_2017b}. Furthermore, the coexistence
of kinetic instabilities with strong plasma turbulence has been
investigated by means of 2D HEB simulations including the effects
of the solar wind expansion~\cite{Hellinger_al_2015, Hellinger_al_2017}.

\subsection{Numerical scheme}

\begin{figure}[t]
\begin{center}
\includegraphics[width=0.6\linewidth]{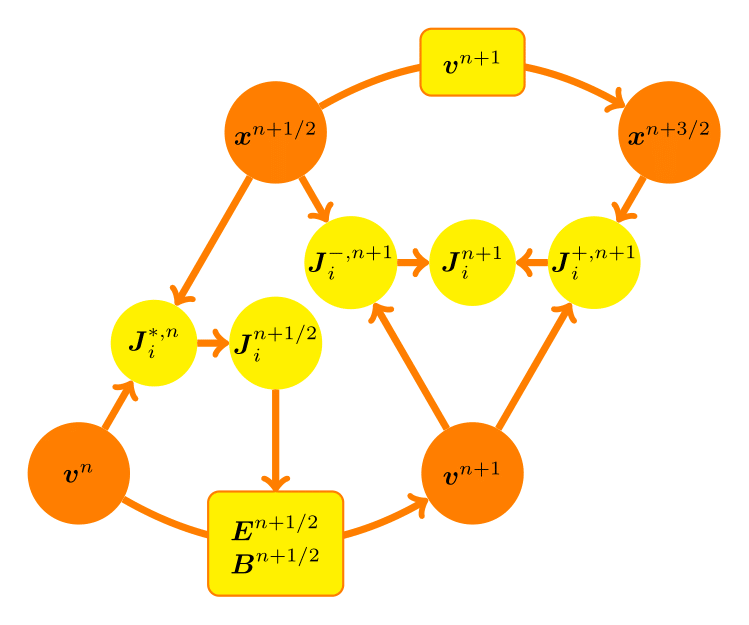}
\caption{Schematic diagram of the code CAMELIA.
{\footnotesize The particle positions and velocities are taken at the different
times, separated by half the (particle) time step $\Delta t/2$, $\vv^n$ and
$\xv^{n+1/2}$, where the subscript $n$ refers to time $t$ and $n+1/2$
to time $t + \Delta t/2$.  At the beginning, there are the current
density, $\Jv_i^{n+1/2} = \Jv(\xv^{n+1/2})$, and the ``free-streaming''
ionic current, $\Jv_i^{\mbox{*},n} = \Jv_i (\xv^{n+1/2},\vv^n)$, as well
as $\Jv^n= \Jv_n(\xv_n)$ and $\Jv_i^n = \Jv_i(\xv_n,\vv_n)$. The
magnetic field is advanced (using two copies via cyclic leapfrog) from
$\Bv_n$ to $\Bv^{n+1/2}$ with $\Ev(\Jv^n,\Jv_i^n,\Bv)$. The Current
Advance Method advances $\Jv_i^{\mbox{*},n}$ to $\Jv_i^{n+1/2}$.
Using $\Bv^{n+1/2}$ and $\Ev^{n+1/2}=
\Ev(\Jv^{n+1/2},\Jv_i^{n+1/2},\Bv^{n+1/2})$, the particles are
advenced using the Boris' scheme to $\vv^{n+1}$ and $\xv^{n+3/2}$ and
moments are collected, $\Jv^{n+3/2} = \Jv(\xv^{n+3/2})$; $\Jv^{n+1}$
is obtained as an average of $\Jv^{n+1/2}$ and $\Jv^{n+3/2}$. Forward
and backward ``free-streaming'' ionic currents, $\Jv_i^{-,n+1} =
\Jv_i(\xv^{n+1/2},\vv^{n+1})$ and $\Jv_i^{+,n+1} =
\Jv_i(\xv^{n+3/2},\vv^{n+1})$, are collected and their average give
$\Jv_i^{n+1}$. Finally, $\Bv^{n+1/2}$ is advanced to $\Bv^{n+1}$.}}
\label{fig:graph}
\end{center}
\end{figure}

CAMELIA is based on the Current Advance Method and Cyclic
Leapfrog (CAM-CL) code of Matthews \cite{Matthews_1994}.  The system
is governed by the Vlasov-fluid equations, comprising the equations of
motions for individual ions, and the electron fluid equations.  The
ions are described by a PIC model, a technique used to solve a certain
class of partial differential equations where particles (or fluid
elements) in a Lagrangian frame are tracked in continuous phase space,
whereas moments of the distribution such as densities and currents are
computed simultaneously on Eulerian (stationary) mesh points. The
central engine of the code is a CAM-CL algorithm, which integrates the
differential equations in a manner that is explicit in time and
spatially local. The plasma has two time-independent components: ion
macro-particles with position and velocities and the magnetic field
specified at the nodes of a regular computing grid. The displacement
current is neglected in Maxwell’s equations, so there is no equation
for the time-evolution of the electric field, which is just a function
of the ion moments (interpolated at grid points from particle data),
the magnetic field, and the electron temperature. (Bi, tri)linear
interpolation is used for moment collection and evaluation of the
Lorentz force at particle positions.  An explicit form of the
time-dependent differential equation is used. A CAM method is used to
advance the ion current density. The original leapfrog particle
advance has been replaced by the more precise Boris'
algorithm~\cite{Boris_1970}, which has been proved to
have an excellent long-term accuracy~\cite{Qin_al_2013}.
This requires the fields to be
known at half time step ahead of the particle velocities, which is
achieved by advancing the current density to this time step, with only
one computational pass through the particle data at each time step
(see Fig.~\ref{fig:graph}). Cyclic leapfrog is used to advance the
magnetic field with a sub-stepping using two copies of the magnetic
field.
\begin{center}
\begin{table}[t]
\centering
\begin{tabularx}{0.95\textwidth}{lccccc}
\br
 & Peak Performance & Clock & Cores/node & RAM/node & Network \\
\mr
SuperMike-II & 146 TFlops & 2.6 GHz & 16 & 32 GB & 40 Gbit/s Infiniband \\ 
Fermi & 2 PFlops & 1.6 GHz & 16 & 16 GB & 5D Torus \\
Cartesius & 1.8 PFlops & 2.6 GHz & 24 & 64 GB & 100 Gbit/s Infiniband \\
Marconi-A1 & 2 PFlops & 2.3 GHz & 36 & 128 GB & Intel OmniPath\\
Marconi-A2 & 11 PFlops & 1.4 GHz & 68 & 16+96 GB & Intel OmniPath \\ 
\br
\end{tabularx}
\caption{Main parameters of the different HPC systems used for scaling tests}
\label{tab:HPCsystems}
\end{table}
\end{center}

\subsection{Performance}

CAMELIA has been optimized and run on many thousands
of cores on High Performance Computing (HPC) systems with different
architectures.  Here we provide scaling tests performed on:
\begin{itemize}
\item SuperMike-II (Intel Sandy Bridge) @ Louisiana State University, Baton Rouge, U.S. (2015)
\item Fermi (IBM Blue Gene/Q) @ CINECA, Bologna, Italy (2016)
\item Cartesius (Intel Haswell) @ SURFsara, Amsterdam, The Netherlands (2017)
\item Marconi-A1 (Intel Broadwell) @ CINECA, Bologna, Italy (2017) 
\item Marconi-A2 (Intel Knights Landing) @ CINECA, Bologna, Italy (2017) 
\end{itemize}
The main features of these HPC systems are summarized and compared in Tab.~\ref{tab:HPCsystems}.

\begin{figure}[t]
\begin{center}
\includegraphics[width=0.48\linewidth]{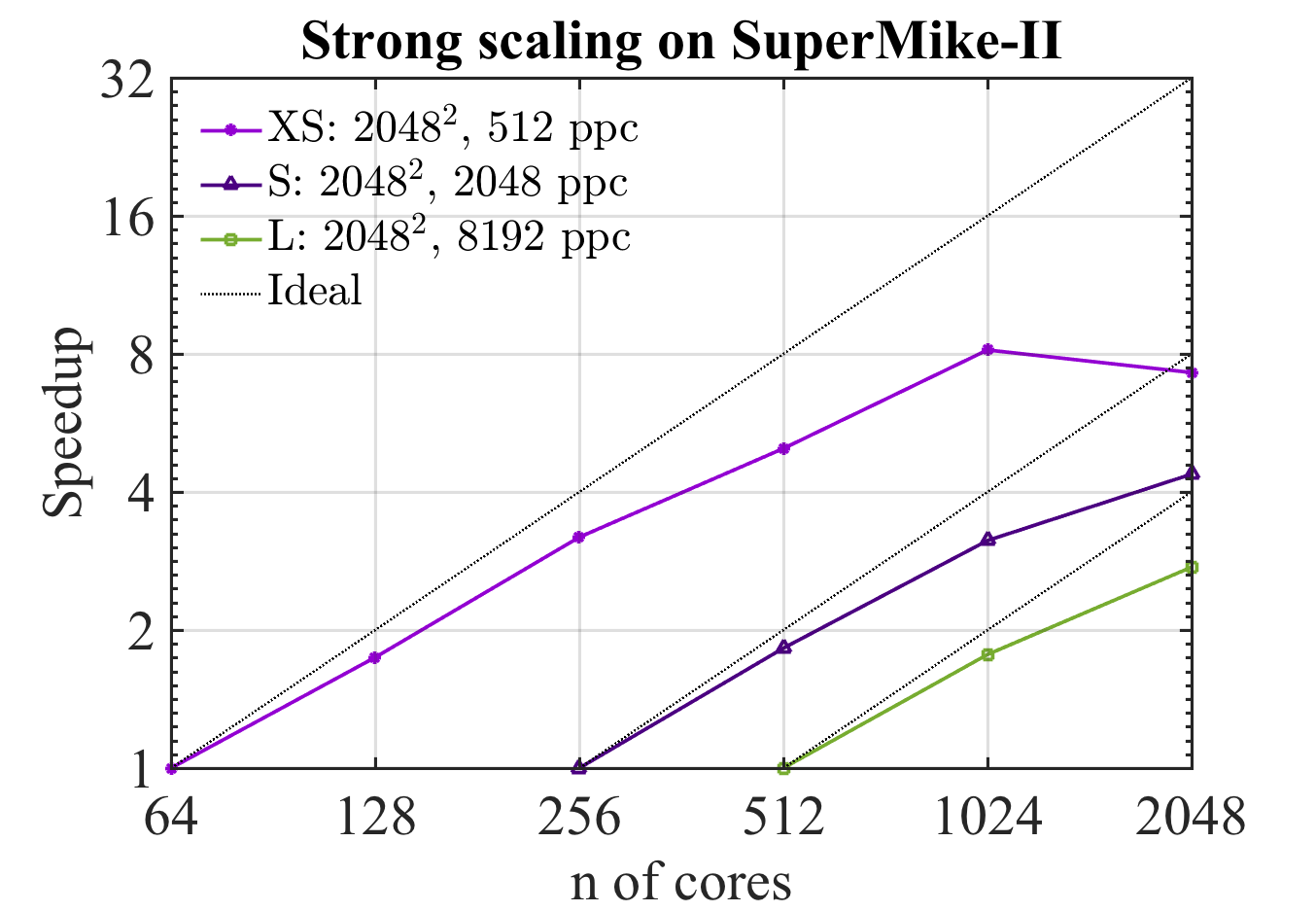}
\includegraphics[width=0.48\linewidth]{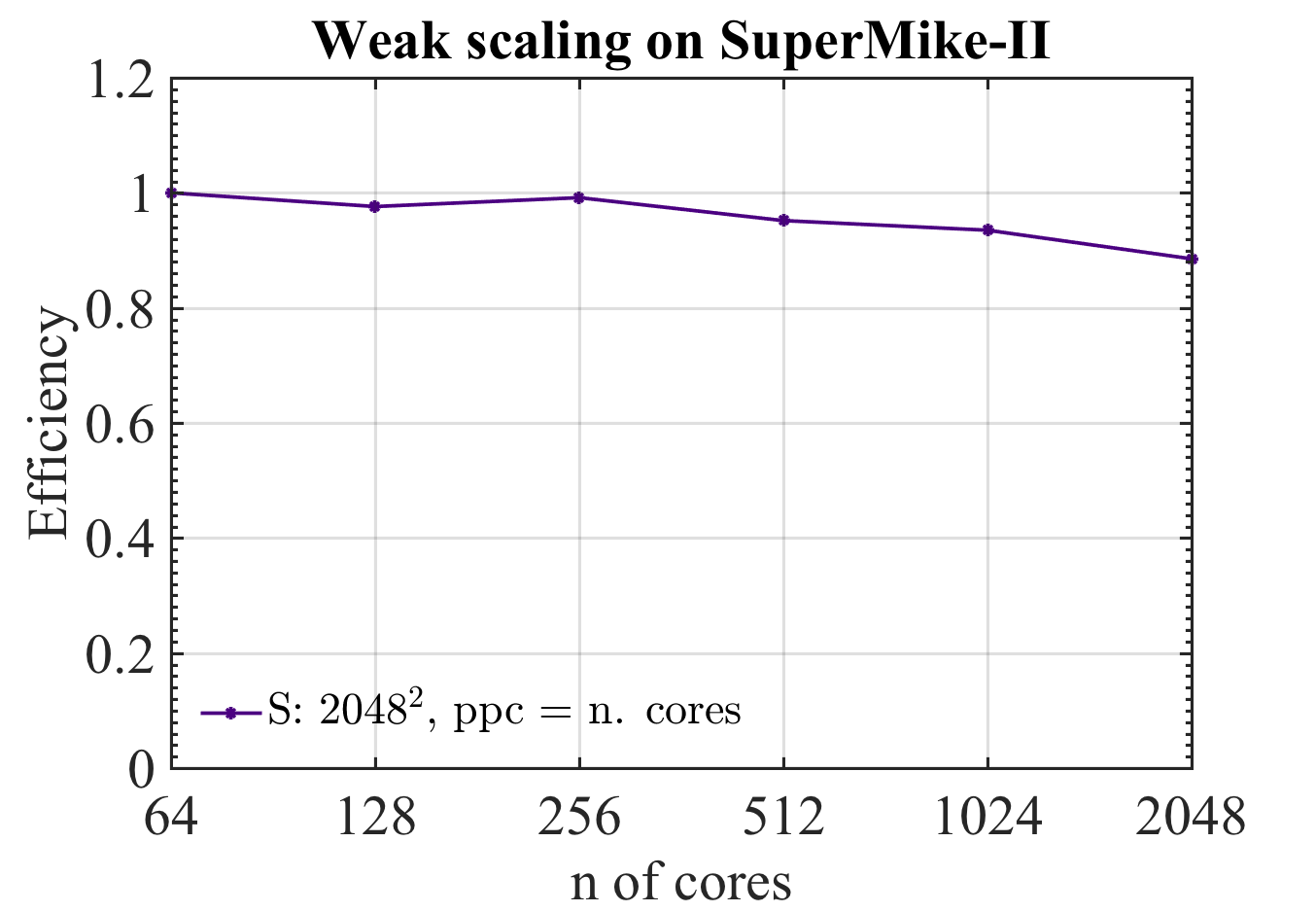}
\caption{Code performances of CAMELIA's pure MPI version on the Sandy
  Bridge architecture, i.e., ``SuperMike-II'': strong scaling (left
  panel) and weak scaling tests (right panel).}
\label{fig:SUPERMIKEII}
\end{center}
\end{figure}

\begin{figure}[t]
\begin{center}
\includegraphics[width=0.48\linewidth]{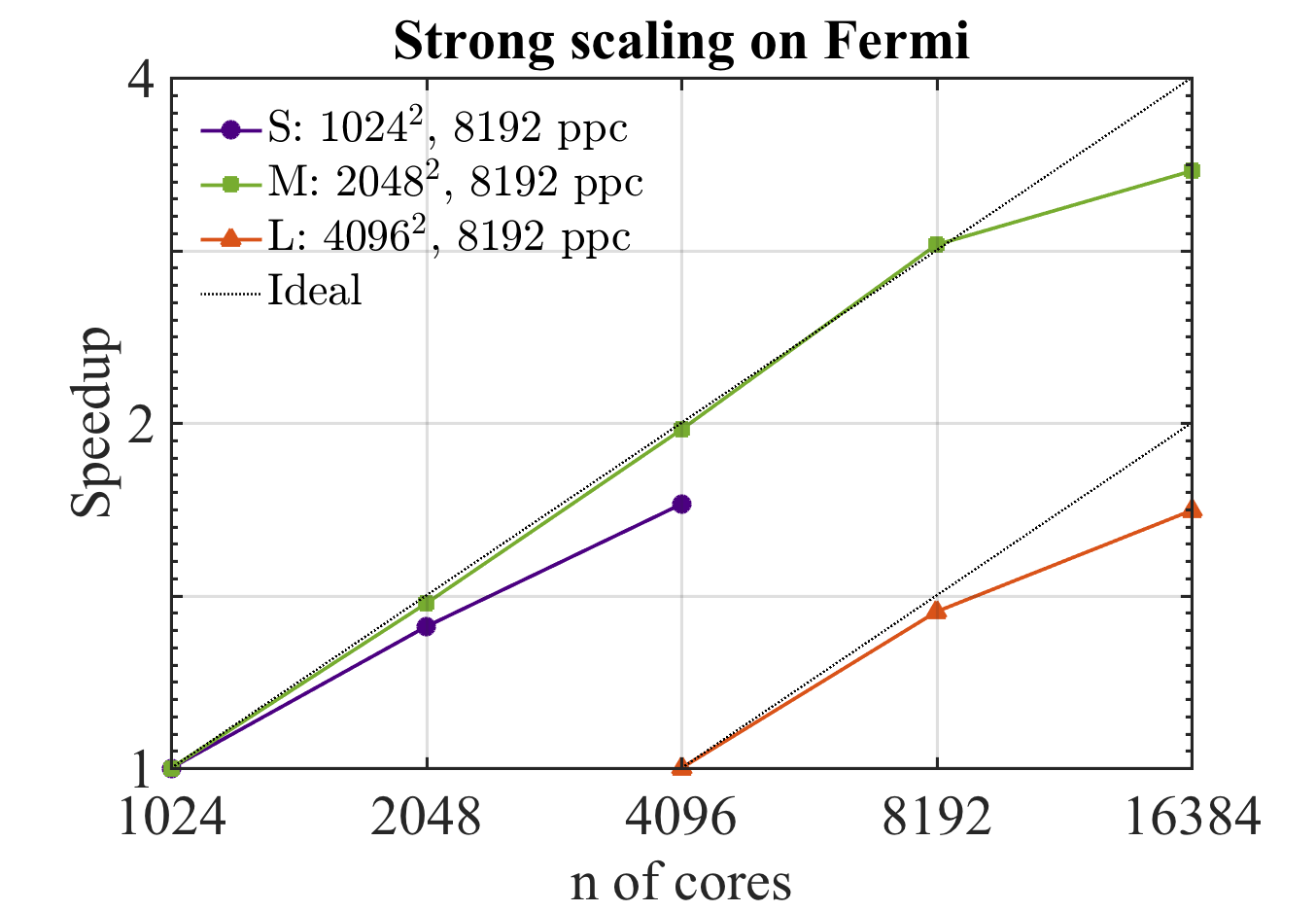}
\includegraphics[width=0.48\linewidth]{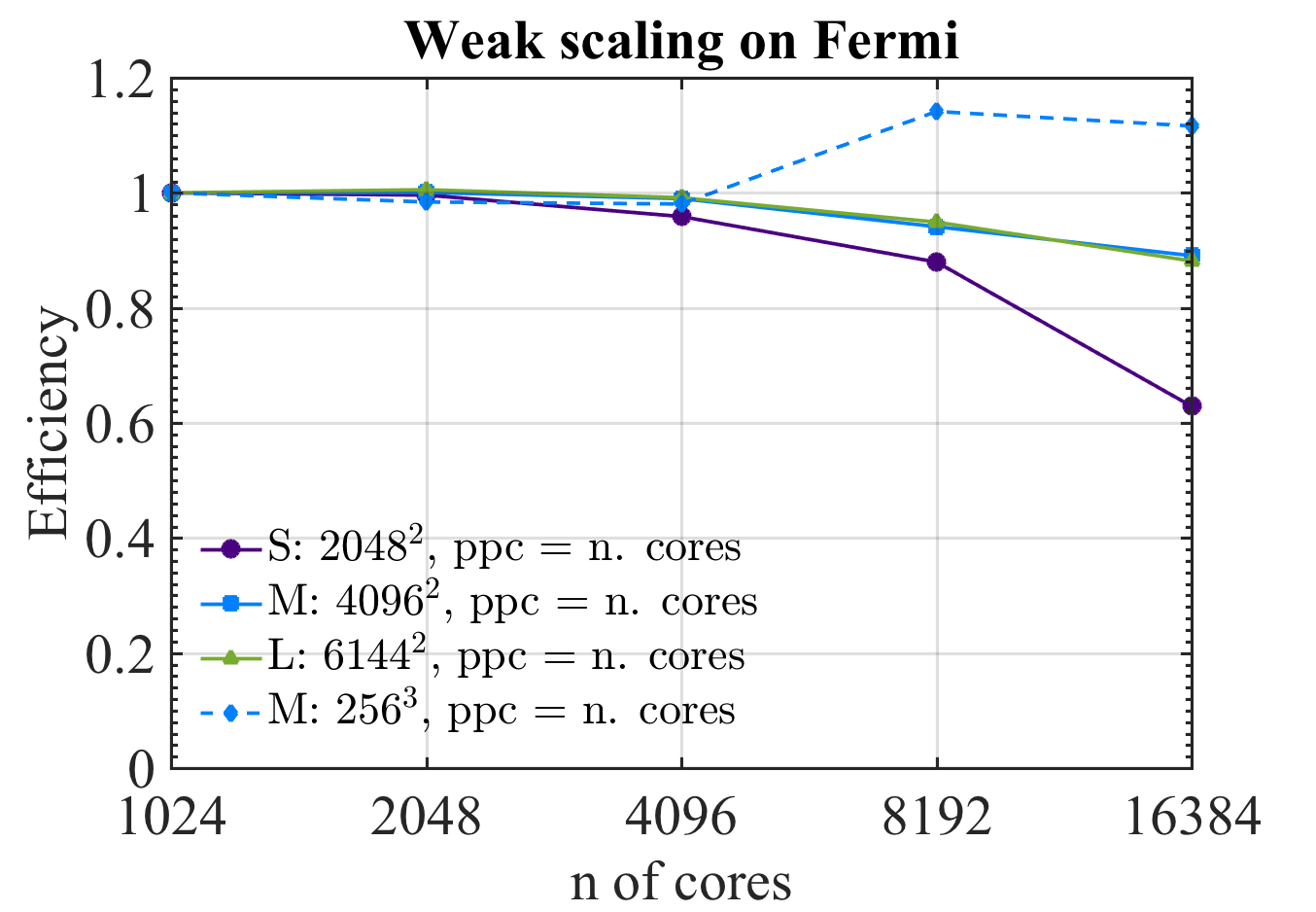}
\caption{The same as in Fig.~\ref{fig:SUPERMIKEII}, but for an IBM
  Blue Gene/Q architecture, i.e., ``Fermi''.}
\label{fig:FERMI}
\end{center}
\end{figure}

\begin{figure}[t]
\begin{center}
\includegraphics[width=0.48\linewidth]{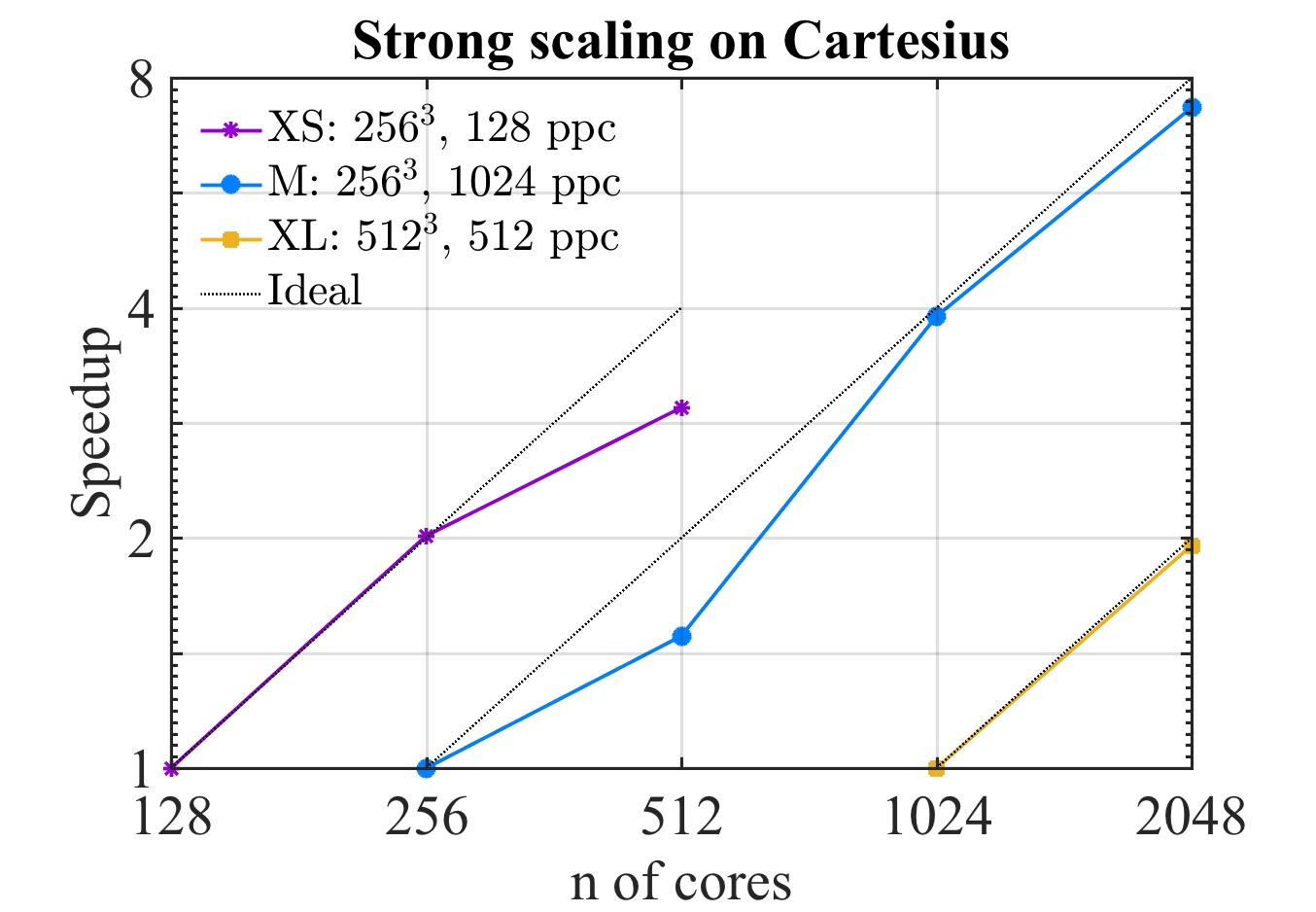}
\includegraphics[width=0.48\linewidth]{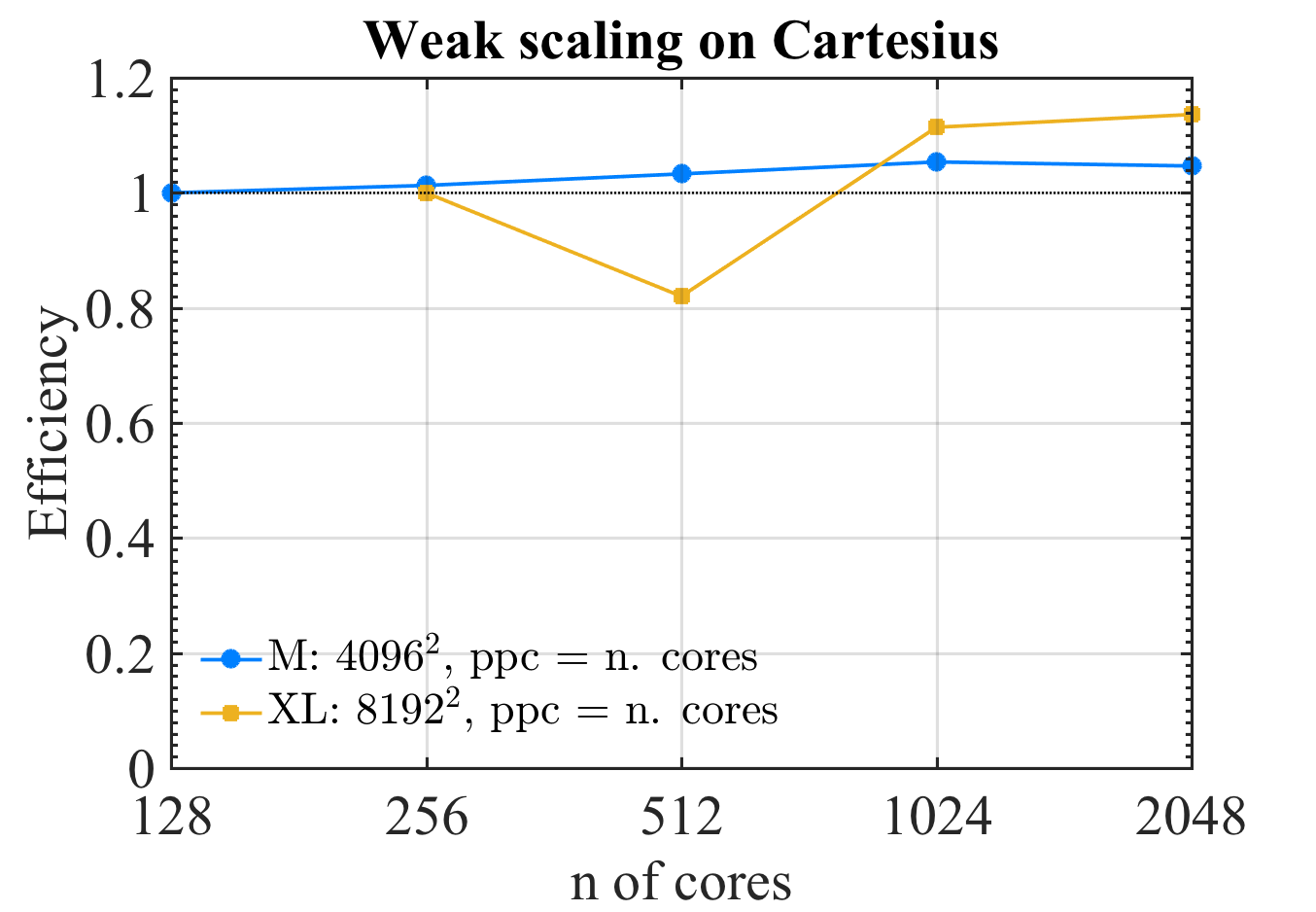}
\caption{The same as in Fig.~\ref{fig:SUPERMIKEII}, but for an 
Intel Hashwell architecture, i.e., ``Cartesius''.}
\label{fig:CARTESIUS}
\end{center}
\end{figure}

\begin{figure}[t]
\begin{center}
\includegraphics[width=0.48\linewidth]{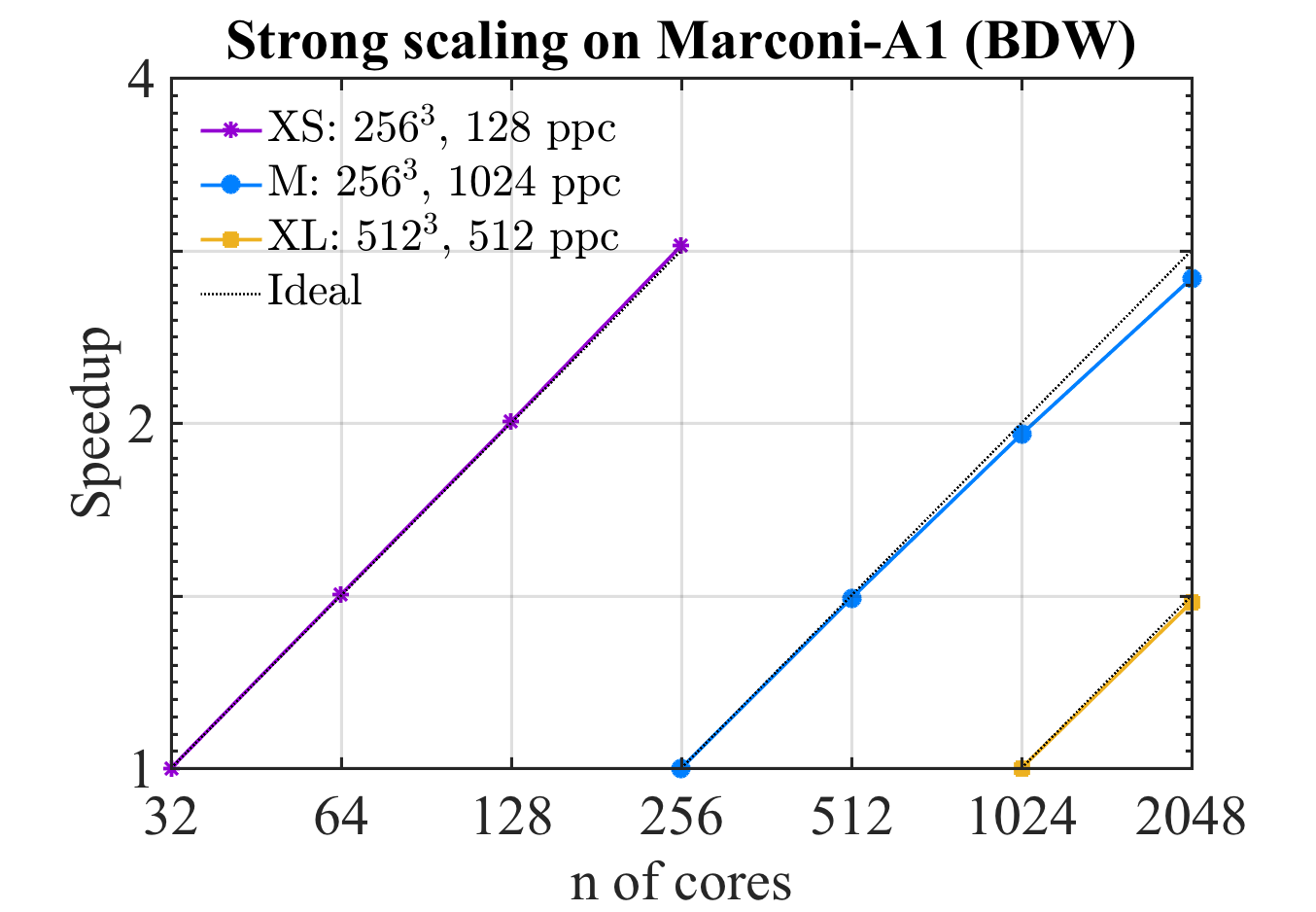}
\includegraphics[width=0.48\linewidth]{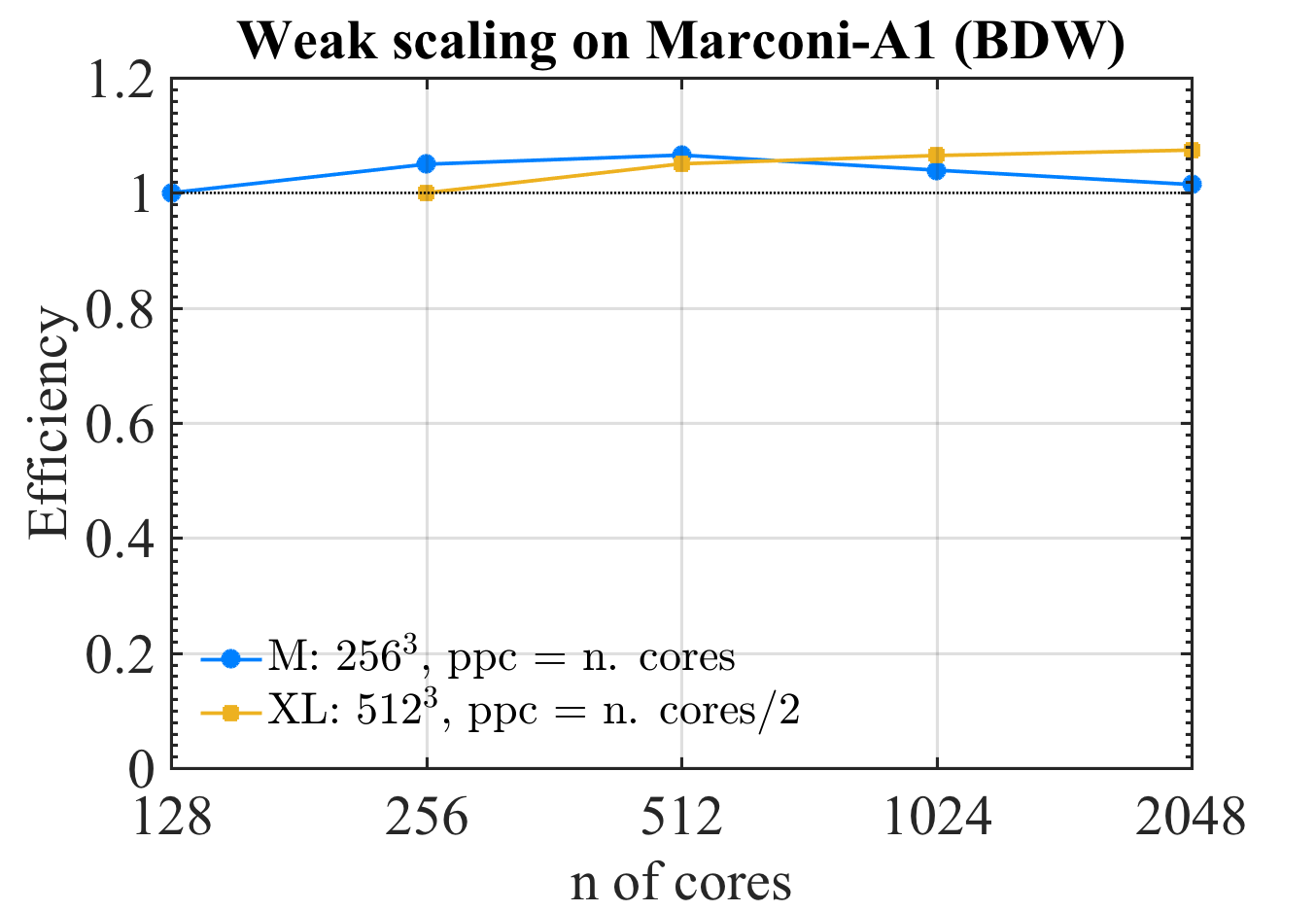}
\caption{The same as in Fig.~\ref{fig:SUPERMIKEII}, but for an Intel
  Broadwell architecture, i.e., ``Marconi-A1 (BDW)''.}
\label{fig:MARCONIA1}
\end{center}
\end{figure}

\begin{figure}[t]
\begin{center}
\includegraphics[width=0.48\linewidth]{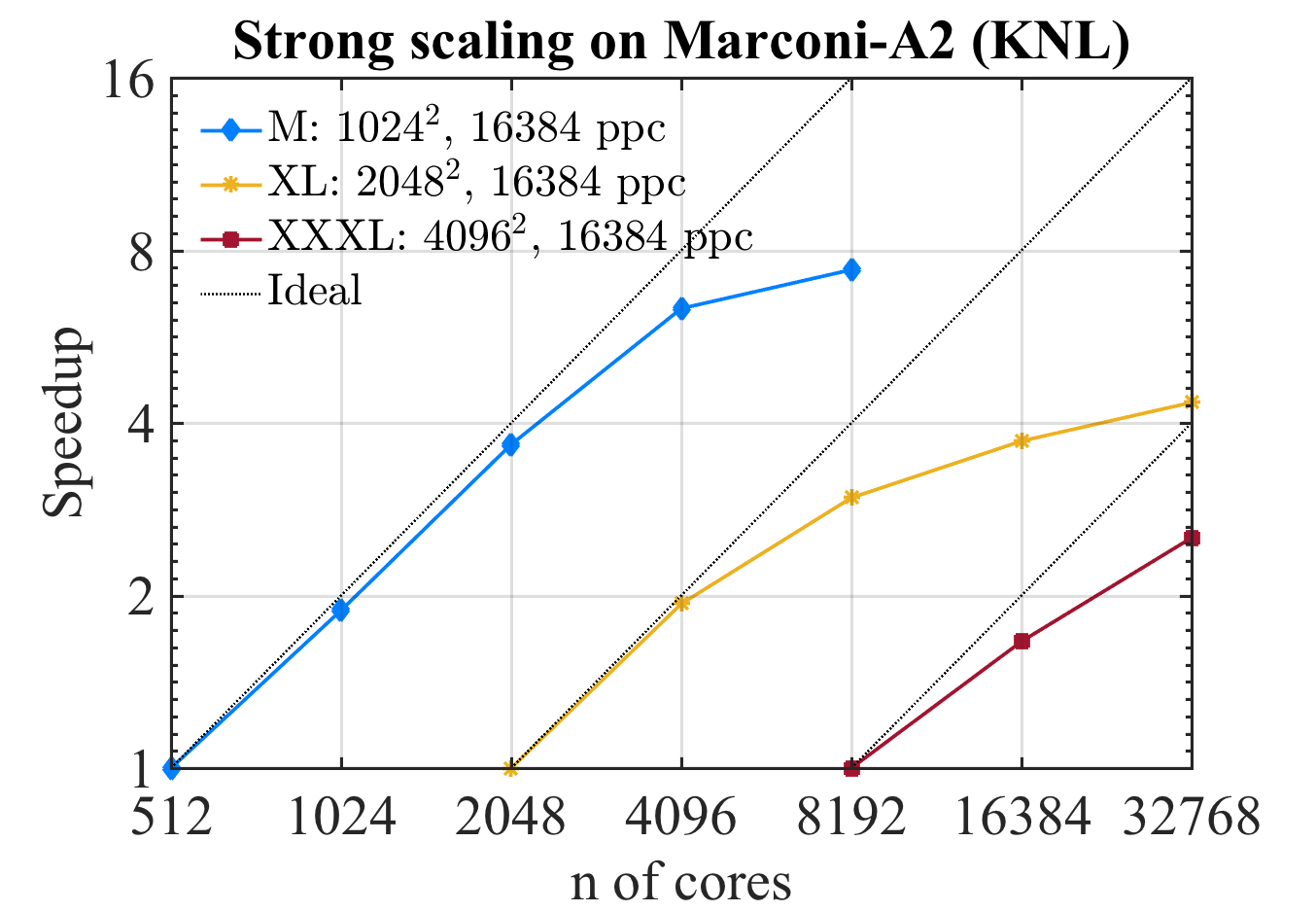}
\includegraphics[width=0.48\linewidth]{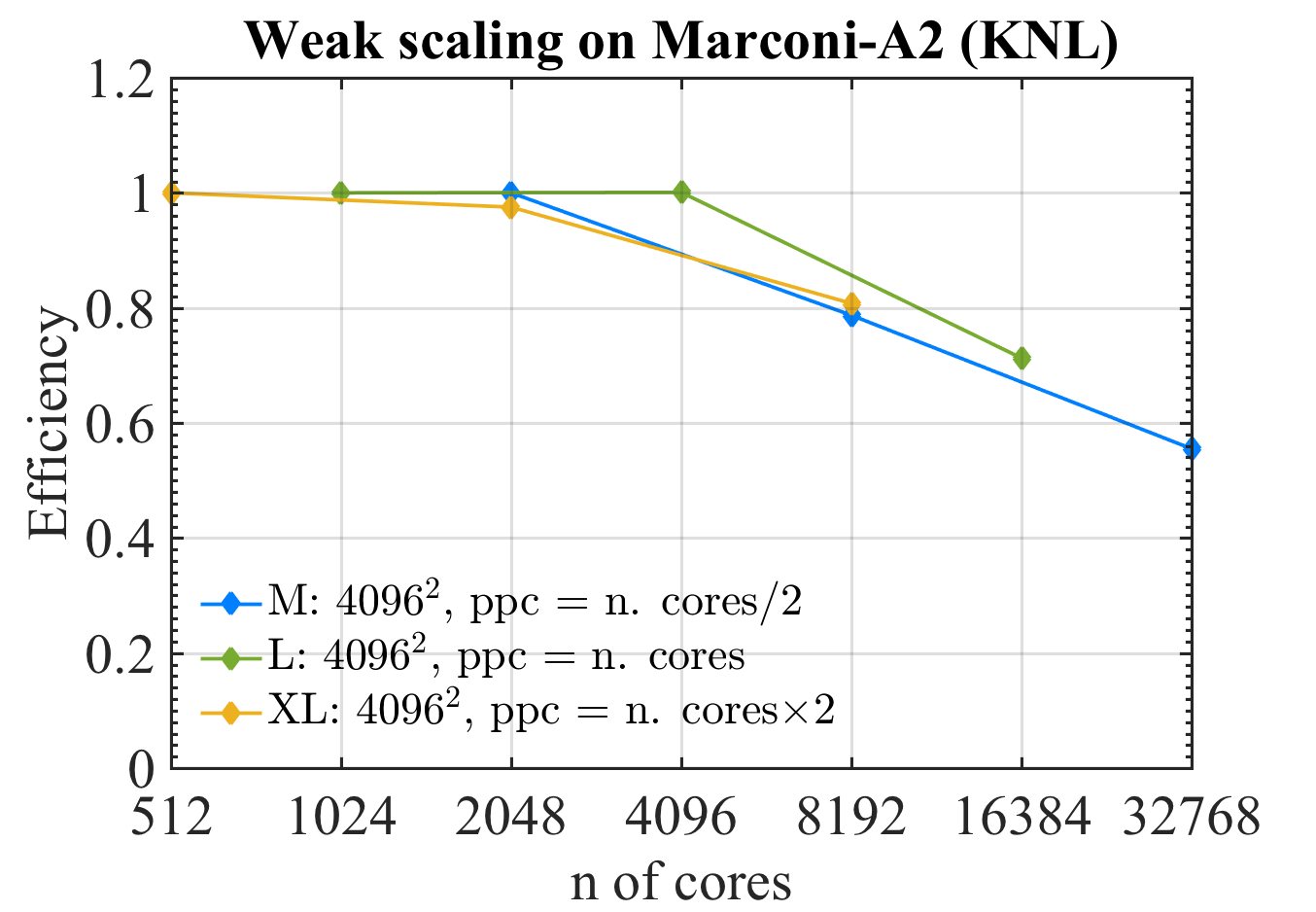}
\caption{Code performances of CAMELIA's hybrid MPI+OpenMP version on
  an Intel Knights Landing architecture, i.e., ``Marconi-A2.}
\label{fig:MARCONIA2}
\end{center}
\end{figure}

For all systems, we performed test simulations with seven
different problem sizes (three for each machine) for measuring 
the strong-scaling performances and four different sizes (one or more 
for each machine) for the weak scaling (see Fig.~\ref{fig:SUPERMIKEII}-
\ref{fig:MARCONIA2}).
In CAMELIA, we evolve  more than 30 different
field components (taking into account both physical variables and
additional temporary auxiliary grid variables), plus the 3 position
components and the 3 velocity components of each particle. Since we
always put thousands of particles in each grid cell, the particles
positions and velocities represent the dominant contribution to the
memory requirement, which is of the order of 4 Bytes (single precision) $\times$ 6
variables $\times$ number of ppc $\times$ number of cells.
In Tab.~\ref{tab:strong}, we list all the different problem sizes,
indicating the total number of particles in the whole simulation
domain and the corresponding total memory requirement. A color has
been associated to each problem size, so that a qualitative comparison 
of code performances for a fixed sized between different machines can 
be easily done by eye. Note that the same global size can result
from different configurations, i.e., number of grid points $\times$ 
number of ppc. These are explicitly indicated in the legend of each
plot.

In Fig.~\ref{fig:SUPERMIKEII}-\ref{fig:MARCONIA2}, we report all the
scaling tests that we performed 
on the above mentioned systems, both for strong scaling (left panels)
and weak scaling (right panels), versus the number of cores.
Until 2016, only a pure MPI parallelization was
implemented, so the number of cores for SuperMike-II, Fermi, Cartesius
and Marconi-A1 also corresponds to the number of MPI processes.
Since early 2017, a hybrid MPI+OpenMP version is also
available, allowing us to fully exploit the Intel Many Integrated Core
Architecture. Test simulations on Marconi-A2 were run with
32 MPI processes per node, i.e., half the number of cores per
node (since 4 over 68 were left available for the operating 
system and I/O operations) and 8 OpenMP threads per MPI task. Therefore,
the code scalability has been measured up to $16384$ MPI tasks $\times \,8$ 
 OpenMP threads/MPI task $= 131072$ total threads.

Fig.~\ref{fig:SUPERMIKEII}-\ref{fig:MARCONIA2} show that both the
strong and the weak scalability are quite good in all the analyzed
architectures. In particular, the weak scalability is very
promising. The parallel efficiency on standard Intel x86 architectures
is near to 1 at least up to 2048 cores.  Both in the Haswell and
Broadwell cases (Cartesius and Marconi-A1) this efficiency is also
slightly greater than 1, and this is probably due to a more optimal
use of the memory with an high number of cores.  Furthermore, the weak
scalability on Blue Gene/Q architectures is also more
promising. Indeed, in almost all the analyzed cases
(with the exception of the smallest one) the parallel efficiency in
near to 1 at least up to 8192 cores.  This increase of parallel
performance is likely due to a combination of factors, the most
important ones being the high speed Torus 5D Blue Gene/Q network and
the ratio between the network (and memory, of course) bandwith and the
CPU clock. The weak scaling tests on F also clearly show that, for
the same problem size, CAMELIA is more efficient in running 3D
simulations than 2D, due to the more favorable ratio between
computations and communications when parallelizing the computational
domain into cubes instead of rectangles.  Finally for what concerns
the parallel performances on KNL machines, the weak parallel
efficiency is close to 1, as expected, up to 4096 cores (i.e. the
same number of nodes as in the Broadwell case), which correspond to
16384 total OpenMP threads, while it is lower for larger number of
cores, mainly due to the fact that a smaller workload per core has
been employed.

\begin{center}
\begin{table}[t]
\centering
\begin{tabularx}{0.6\textwidth}{YYYY}
\br
Size & Particles                  & RAM    & Color  \\ 
\mr
XS   & $\sim 2.1 \times 10^{9} $  &  48 GB & \textcolor{violet}{violet} \\
S    & $\sim 8.6 \times 10^{9} $  & 192 GB & \textcolor{indigo}{indigo} \\
M    & $\sim 1.7 \times 10^{10}$  & 384 GB & \textcolor{blue}{blue} \\
L    & $\sim 3.4 \times 10^{10}$  & 768 GB & \textcolor{green}{green} \\
XL   & $\sim 6.7 \times 10^{10}$  & 1.5 TB & \textcolor{yellow}{yellow} \\
XXL  & $\sim 1.4 \times 10^{11} $ & 3.0 TB & \textcolor{orange}{orange} \\
XXXL & $\sim 2.7 \times 10^{11} $ & 6.0 TB & \textcolor{red}{red} \\
\br
\end{tabularx}
\caption{Problem sizes for strong scaling tests}
\label{tab:strong}
\end{table}
\end{center}

\begin{center}
\begin{table}[t]
\centering
\begin{tabularx}{0.6\textwidth}{YYYY}
\br
Size & Particles/MPI            & RAM/MPI    & Color  \\ 
\mr
S    & $\sim 4.2 \times 10^{7} $  &  96 MB & \textcolor{indigo}{indigo} \\
M    & $\sim 1.7 \times 10^{7}$  & 384 MB & \textcolor{blue}{blue} \\
L    & $\sim 3.4 \times 10^{7}$  & 768 MB & \textcolor{green}{green} \\
XL   & $\sim 6.7 \times 10^{7}$  & 1.5 GB & \textcolor{yellow}{yellow} \\
\br
\end{tabularx}
\caption{Problem sizes for weak scaling tests}
\label{tab:weak}
\end{table}
\end{center}


\section{Conclusions}
\label{sec:conclusions}

CAMELIA is a 3D hybrid kinetic numerical code, suitable for
investigating the development of plasma turbulence and its interplay
with kinetic instablities over a wide range of scales. Indeed, it allows for
the simultaneous modelling of the MHD and the sub-ion-scale dynamics,
fully capturing the transition between the two regimes.  Being highly
performing on many different HPC systems, CAMELIA allows us to perform
state-of-the-art 2D HPIC simulations in a walltime of a few hours and
3D HPIC simulations within a day. In the last few years, the results 
of the HPIC simulations performed with the code CAMILIA showed
a remarkable agreement with solar
wind observations, both in 2D and 3D.

As far as 3D simulations are concerned, we showed that a $256^3$
compuational domain with a spatial resolution of $\di/4$ and $512$
particle-per-cell may be accurate enough for investigating the
development of the turbulent cascade simultaneously from MHD to
sub-ion scales in an intermediate- or small-beta plasma. This
``reduced'' setting can therefore be safely employed for future
parameter studies and convergence studies in three
dimensions. However, based on \cite{Franci_al_2015b,Franci_al_2016b},
we can reasonably expect that a larger number of particles would be
mandatory for properly investigating the evolution of the proton
temperature, or the spectral properties in systems with $\beta \gtrsim
1$.

\section{Acknowledgments} 

The authors thank T. Tullio for useful comments and suggestions.
LF is funded by Fondazione
Cassa di Risparmio di Firenze through the project ``Giovani
Ricercatori Protagonisti''. PH acknowledges GACR grant 15-10057S. 
CHKC is supported by an STFC Ernest Rutherford Fellowship.
We acknowledge PRACE for awarding us access to resource Cartesius based
in the Netherlands at SURFsara through the DECI-13 (Distributed
European Computing Initiative) call (project HybTurb3D), and the
CINECA award under the ISCRA initiative, for the availability of high
performance computing resources and support (grants HP10C877C4,
HP10CVCUF1, HP10B2DRR4, HP10BUUOJM). We warmly thank Frank L\"offler
for providing HPC resources through the Louisiana State University
(allocation hpc hyrel14). Data deposit was provided by the EUDAT
infrastructure (https://www.eudat.eu/).

\section*{References}
\bibliographystyle{iopart-num}

\end{document}